\documentclass[useAMS,usenatbib,twocolumn]{mnras}
\usepackage{natbib}
\usepackage{graphicx}
\usepackage{times}
\usepackage{rotate}
\usepackage{hyperref}
\usepackage{booktabs}
\usepackage{multirow}
\usepackage{url} 
\usepackage[dvipsnames]{xcolor}
\usepackage{ amssymb }
\usepackage{newtxtext,newtxmath}
\usepackage[normalem]{ulem}

\newcommand{\redm}{redMaPPer }

\newcommand{\cosmolike}{\fontfamily{lmtt}\selectfont CosmoLike}

\newcommand{\rmd}{{\rm d}}

\newcommand{\ds}{\ensuremath{\Delta\Sigma}}
\newcommand{\wpp}{w_\mathrm{p}}
\newcommand{\nbar}{\bar{n}_\mathrm{gal}}
\newcommand{\sbgs}{\large\mathcal{S}_\mathrm{BGS}}
\newcommand{\sfull}{\large\mathcal{S}_{M_{\ast [50, 100]}, \mathrm{full}}}
\newcommand{\sbin}{\large\mathcal{S}_{M_{\ast [50, 100]}, \mathrm{bins}}}

\newcommand{\h}{\ensuremath{h^{-1}}~}

\def\rmd{{\rm d}}

\def\rmm{{\rm m}}

\def\rms{{\rm s}}

\def\rms{{\rm s}}

\def\omm{{\Omega_\mathrm{m}}}
\def\s8{{\sigma_8}}
\def\m50100{M_{\ast, [50, 100]}}
\def\redm{\texttt{redMaPPer}}
\def\camira{\texttt{CAMIRA}}

\defcitealias{huang2021outer}{H+22}

\begin{document}
  

\title[Cluster Cosmology Without Cluster Finding]{Cluster Cosmology Without Cluster Finding}





\author[Xhakaj et al.]{Enia Xhakaj$^{1}$\thanks{E-mail: exhakaj@ucsc.edu}, Alexie Leauthaud$^{1}$, Johannes Lange$^{1,2,3,4}$,  Elisabeth Krause$^{5}$, Andrew Hearin$^{6}$, 
\newauthor
 Song Huang$^{7}$, Risa H. Wechsler$^{2}$, Sven Heydenreich$^{1}$
\\
\\
$^{1}$Department of Astronomy and Astrophysics, University of California, Santa Cruz, 1156 High Street, Santa Cruz, CA 95064 USA \\
$^2$Kavli Institute for Particle Astrophysics and Cosmology and Department of Physics, Stanford University, CA 94305, USA\\
$^3$Department of Physics, University of Michigan, Ann Arbor, MI 48109, USA\\
$^4$Leinweber Center for Theoretical Physics, University of Michigan, Ann Arbor, MI 48109, USA\\
$^5$ Department of Astronomy/Steward Observatory University of Arizona, 933 North Cherry Avenue, Tucson, AZ 85721-0065, USA\\
$^{6}$High-Energy Physics Division, Argonne National Laboratory, Argonne, IL 60439, USA \\
$^{7}$ Department of Astronomy, Tsinghua University, Beijing, China, 100084 \\
}

\maketitle
\label{firstpage}

 
\begin{abstract} 
We propose that observations of super-massive galaxies contain cosmological constraining power similar to conventional cluster cosmology, and we provide promising indications that the associated systematic errors are comparably easier to control. We consider a fiducial spectroscopic and stellar mass complete sample of galaxies drawn from the Dark Energy Spectroscopic Survey (DESI) and forecast how constraints on $\omm$--$\s8$ from this sample will compare with those from number counts of clusters based on richness $\lambda.$ At fixed number density, we find that massive galaxies offer similar constraints to galaxy clusters. However, a mass-complete galaxy sample from DESI has the potential to probe lower halo masses than standard optical cluster samples (which are typically limited to $\lambda\gtrsim20$ and $M_\mathrm{halo} \gtrsim 10^{13.5} \mathrm{M}_\odot/h$); 
additionally, it is straightforward to cleanly measure projected galaxy clustering $\wpp$ for such a DESI sample, which we show can substantially improve the constraining power on $\omm.$ We also compare the constraining power of $M_*$-limited samples to those from larger but mass-incomplete samples (e.g., the DESI Bright Galaxy Survey, BGS, Sample); relative to a lower number density $M_*$-limited samples, we find that a BGS-like sample improves statistical constraints by 60\% for $\omm$ and 40\% for $\s8$, but this uses small scale information which will be harder to model for BGS. Our initial assessment of the systematics associated with super-massive galaxy cosmology yields promising results. The proposed samples have a $\sim$ 10\% satellite fraction, but we show that cosmological constraints may be robust to the impact of satellites. These findings motivate future work to realize the potential of super-massive galaxies to probe lower halo masses than richness-based clusters and to avoid persistent systematics associated with optical cluster finding.
\end{abstract}
 

\section{Introduction}

The current $\Lambda$CDM paradigm successfully explains numerous cosmological phenomena, including the formation and growth of large-scale structure and the cosmic expansion of the Universe \citep[e.g.,][]{expans_13, expans_14}. However, the physical origin of cosmic acceleration is still poorly understood. Cosmic acceleration is studied through probes such as Type Ia Supernovae (SNe), baryon acoustic oscillations (BAO), weak gravitational lensing, and galaxy clusters \citep[e.g.][and references therein]{albrecht_06, dawson_18, Abdalla2022_JHEAp_34_49}. Type Ia Supernovae, and baryon acoustic oscillations probe the expansion history, while weak gravitational lensing and galaxy clusters probe the growth of dark matter structure. In turn, the expansion history of the Universe and the growth of dark matter structure shed light on cosmic expansion.

Galaxy clusters are formed from the highest peaks of the matter density field. Their abundance and spatial distribution contain a wealth of information on the growth of dark matter structure \citep{schuecker_03, schmidt_04, Bonamente_06, albrecht_06, allen_08, ettori_09, chandra_09, henry_09,mantz_10, rozo_10, allen_11,  expans_13, Mantz2015_MNRAS_446_2205, PlanckCollaboration2016_AA_594_24, Bocquet2019_ApJ_878_55, Costanzi_21, to_21, chiu_22, salvati_22, lesci_23, park_23}. Recently, optical cluster counts and weak gravitational lensing have offered competitive constraints on $\s8$ and $\omm$ \citep[e.g.][]{rozo_10, desi20_clusters, to_21, Costanzi_21, Giocoli_21, Lesci_22, park_23}. Optical surveys such as the Dark Energy Survey \citep[DES,][]{DES_05} and Hyper Suprime-Cam \citep[HSC,][]{Aihara2018FirstProgram} have detected larger cluster samples than surveys in other wavelengths providing better statistics for cosmological analyses. Three important systematics associated with optical cluster finding are selection effects, projection effects (the failure to separate halos along the line of sight into distinct objects), and cluster miscentering. These effects can bias cosmology if not modeled correctly \citep[e.g.][]{Erickson2011, desy1miscentering, costanzi_19_systematics, sunayama_2020S, wu_12}. A great deal of effort has been devoted to building mocks that study projection effects \citep{DeRose2019, Korytov2019, sunayama_2020S, joe_22, wu_22, Wechsler2022, park_23}. However, mocks with accurate red galaxy populations have proven challenging to build. In addition to galaxy color, the spatial distribution of cluster satellites is also challenging to model \citep{DeRose2019, Korytov2019, joe_22, Wechsler2022, to_2023_cardinal}. 

Another popular method to constrain the growth of structure is the combination of galaxy--galaxy lensing and projected clustering. Galaxy--galaxy lensing is proportional to $b \omm \s8^2$ at large radial scales, where $b$ is the galaxy bias \citep[e.g.,][]{yoo_06}, $\omm$ is the mean matter density of the Universe, and $\s8$ is the amplitude of the power spectrum. Projected clustering, on the other hand, scales as $b^2 \s8^2$. When fit jointly, the bias term is constrained, allowing for a measurement of $\omm \s8$  \citep{Baldauf_2010, more_2013, Mandelbaum2013, Cacciato2013_MNRAS_430_767, more_15, Leauthaud2017_MNRAS_467_3024, Lange2019_MNRAS_488_5771, singh2020}. Cosmological constraints from lensing and clustering analyses have often been carried out using red galaxy samples. For example, the Baryon Oscillation Spectroscopic Survey \citep[BOSS,][]{dawson_13} provided a large number of massive galaxies with spectroscopic redshifts that have been used for lensing plus clustering studies \citep{Miyatake2015_ApJ_806_1, Leauthaud2017_MNRAS_467_3024, singh2020, sunayama_2020S, Lange2022_MNRAS_509_1779, troster_22, Leauthaud2022_MNRAS_510_6150, amon22, lange_2023}.  The DES survey instead used photometric samples of red galaxies \citep{desi19, desi22}. However, these galaxy samples have typically been selected with color cuts and are rarely complete, except at the highest galaxy masses \citep{leauthaud_16}. As such, unlike studies of galaxy clusters where abundance is a key constraint, traditional lensing plus clustering analyses do not use the observed number density for cosmological constraints\footnote{For example, the $f_\Gamma$ parameter in \citet{Lange2022_MNRAS_509_1779} allows the amplitude of the central halo occupation float as a free parameter which removes any constraining power from the observed number density.}.

Here we introduce the idea of using mass-complete samples of super-massive galaxies to constrain cosmology by applying a methodology that unifies aspects of cluster cosmology and traditional lensing plus clustering analyses. Table \ref{tab:intro_table} summarizes the key observational probes typically used for the two classic approaches. As highlighted in Table \ref{tab:intro_table}, cluster cosmology has typically not used information from the clustering of clusters (but see \citealt{wu_12, park_21, to_21})  whereas lensing plus clustering typically does not use number density as a constraint. As explained in the previous paragraph, this is because galaxies samples from surveys such as BOSS are highly incomplete. However, the Dark Energy Spectroscopic Survey \citep[DESI,][]{desi_16} will change this picture \citep{desi_16}. Indeed, DESI will be deep enough to detect large samples of massive galaxies which are complete above $M_\ast=11.5$ $\mathrm{M_\odot}$ at intermediate redshifts (e.g., $z<0.6$). We propose that mass complete samples from DESI can be used in a similar fashion to traditional cluster cosmology but with two added advantages: 1)it may be possible to bypass some of the complications associated with cluster finding (see below), and 2) it will be possible to take advantage of spectroscopic galaxy clustering measurements from DESI. 

New results from \citet[][hereafter \citetalias{huang2021outer}]{huang2021outer} show that super-massive galaxies can be used to trace dark matter halos with scatter comparable to state-of-the-art red sequence methods. This work suggests that the stellar mass measured within $R = [50,100]~\mathrm{kpc}$ (the outer envelope of the stellar mass distribution -- hereafter the {\it outskirt mass}) yields a halo mass tracer with scatter similar to red-sequence-based cluster finders such as \redm{} (e.g.,\citealt{Rykoff2014, Rozo2014, Rozo2015a, Rozo2015b, Rykoff2016}) and \camira{} (e.g., \citealt{Oguri2014, Oguri2018}). The outskirt mass has already been measured in multiple surveys \citep{li_21} and will be well measured with future surveys like the Rubin Observatory Legacy Survey of Space and Time \citep[LSST,][]{Ivezic2008LargeDesign}. 

Given the results of \citetalias{huang2021outer}, we explore here the idea that massive galaxies from DESI can be used to trace halos, thus bypassing the cluster-finding process. This method will not rely on prior knowledge about red galaxies in clusters and will not suffer from cluster selection effects such as those described in \citet{desi20_clusters}. However, we expect this method to have other systematics that must be studied further. In particular, about $10-20\%$ of massive galaxies will be satellite galaxies. For example, \citet{Mandelbaum_06} predict $f_\mathrm{sat} = 0.16 \pm 0.09$ for early type galaxies of mass $M_\ast=10^{11.3} \mathrm{M}_\odot$. \citet{Reddick_13} find $f_\mathrm{sat} = 0.13 \pm 0.05$ for galaxies of mass $M_\ast=10^{11.6} \mathrm{M}_\odot$. Finally, \citet{saito_16} find $f_\mathrm{sat} = 0.11$ for CMASS galaxies at $z = 0.55$ and of mass $M_\ast>10^{11.6} \mathrm{M}_\odot$. In order to use our proposed methodology, it will be important to model the impact of satellite galaxies. However, satellite modeling may be more straightforward than understanding and making mocks of red galaxy populations. 
This paper aims to study how super-massive galaxies from the DESI survey could be used for cosmological constraints and how this methodology compares with other traditional analyses (see Table \ref{tab:intro_table}). Our main probes are galaxy--galaxy lensing ($\ds$), projected galaxy clustering ($\wpp$), and galaxy number density ($\nbar$). We adopt three narrow mass bins in the outer stellar mass range\footnote{Note that this uses {\it outskirt} stellar mass rather than {\it total} stellar mass. This selection probes host halos of mass $\sim 10^{13.5}\mathrm{M}_\odot/h$.} $10^{10.8}-10^{12.1} \mathrm{M_\odot}$ and a redshift bin of $z \in [0.3, 0.6]$. We assume the full $1000~\mathrm{deg^2}$ area of HSC. We show how our constraints compare with traditional methods such as cluster cosmology and lensing plus clustering analyses. It is important to note that because this is a new method, systematics (e.g., satellite modeling) must be quantified and understood. This will be the goal of future work. Here we focus first on the \emph{relative statistical constraining power of different techniques and analysis choices}. 

This paper is structured as follows. In Section \ref{theory}, we introduce our model. In Sections \ref{method} and \ref{errors}, we introduce our model fitting routine and the derivation of the covariance matrix. We present our results in Section \ref{sec:results}. We discuss our results and possible future directions in Section \ref{sec:discussion}. Finally, we summarize and conclude in Section \ref{conclusions}. We adopt the cosmology of the Covariance AbacusSummit Boxes \citep{Maksimova_2021}, namely, a flat, $\Lambda \mathrm{CDM}$ cosmology with  $\Omega_\mathrm{\rm m} = 0.307$, $\Omega_\mathrm{\rm b} = 0.0482$, $\sigma_8 = 0.829$, $h = 0.678$, $n_\mathrm{\rm s}=0.9611$, corresponding to the best-fit Planck cosmology \citep{planck_18}.

\setlength{\tabcolsep}{20pt}
\renewcommand{\arraystretch}{1.3}

\begin{table*}
\caption{Typical observables and analysis choices for cluster cosmology and lensing plus clustering. Cluster cosmology uses $\ds$ and cluster abundance (which offers the same information as $\nbar$) but typically does not use the clustering of clusters (except in more recent studies such as \citealt{mana_13, park_21, to_21}). Cluster samples are also studied in bins of cluster richness (for example, $\lambda$), defined as the number of red galaxies associated with a cluster within a certain radius. The traditional lensing plus clustering technique uses $\ds$ and $\wpp$ (but not $\nbar$) and often uses one large sample without any binning in galaxy mass. Here we propose to unify the two techniques using complete samples of super-massive galaxies from the DESI survey. In this approach, all observables are used as constraints, and the data can be divided into mass bins.}
\begin{tabular}{l|cccc}
                                        & \textbf{$\ds$} & \textbf{$\wpp$} & \textbf{$\nbar$} & Mass Bins \\ \hline
\textbf{Clusters}                       & $\checkmark$   &                 & $\checkmark$     & $\checkmark$       \\ \hline
\textbf{Traditional Lensing+Clustering} & $\checkmark$   & $\checkmark$    &                  &                    \\ \hline
\textbf{Super-Massive Galaxies}         & $\checkmark$   & $\checkmark$    & $\checkmark$     & $\checkmark$       \\ \hline
\end{tabular}
\label{tab:intro_table}
\end{table*}
\section{Theoretical Framework}\label{theory}

Our goal is to study the cosmological constraining power of complete samples of super-massive galaxies. Our model is based on the halo model in which all dark matter is clumped into spherically over-dense regions known as halos. The following sections describe how the data vectors are modeled. We first introduce our mass to observable relations. We then describe how we model the data vector.

\subsection{The Mass to Observable Relation}\label{sec:mass_to_observable}

\subsubsection{Philosophy}

Traditionally, the correlation between halo mass and stellar mass has been modeled through the stellar-halo mass relation  (SHMR, e.g., \citealt{behroozi_2010, leauthaud_2012, Tinker2017, Kravtsov2018}; also see \citealt{Wechsler2018} for a recent review). This relation has generally been calibrated using the measured stellar masses of central galaxies. The total stellar mass of galaxies can be divided into inner and outer components. Recently, \citet{h20, huang2021outer} showed that the inner light of massive galaxies has a very poor correlation with halo mass (scatter of halo mass at fixed stellar mass, $\sigma_\mathrm{log_{10}{M_h}|log_{10}{M_*}}$, is about 0.8 dex), unlike the outskirt mass of massive galaxies ($\sigma_\mathrm{log_{10}{M_h}|log_{10}{M_*}} \sim 0.4$ dex)\footnote{Note: The methodology used in \citetalias{huang2021outer}  cannot constrain the scatter of the SHMR. Instead, \citetalias{huang2021outer} estimate the scatter of halo mass within a fixed number density bin. Given the slope of the SHMR and the bin width, this value should always be higher than the scatter value of the SHMR defined in a more traditional sense. Please read \citetalias{huang2021outer} for more details.}. The latter is comparable to state-of-the-art red sequence cluster finders and may outperform red sequence methods below $\lambda=10$ (Figure 8, \citetalias{huang2021outer}). 

For this reason, instead of using total galaxy mass, we adopt the {\it outskirt mass}. More specifically, this is the stellar mass measured at $R = 50-100 \mathrm{kpc}$ (hereafter $M_{\ast, [50, 100]}$). Although an SHMR using $M_{\ast, [50, 100]}$ has not yet been calibrated, this paper assumes an unbroken power-law relation with log-normal scatter. This is justified because we are focusing on just the very high mass range ($\mathrm{log_{10}}\m50100 \in [10.8, 12.1]$) well above the pivot mass \citep{leauthaud_2012} where the SHMR can be approximated as a simple log-normal relation. Furthermore, we assume that our mass-to-observable relation models only central galaxies and neglects satellites. The impact of satellites will be discussed in Section \ref{sec:satellites}.

Our log-normal model ($\mathcal{N}$) consists of three free parameters: the slope of the power law ($\beta$), the scatter in stellar mass or richness ($\sigma_{\mathrm{log_{10}} M_\ast}$ or $\sigma_{\mathrm{log_{10}} \lambda}$), and the amplitude of the stellar mass function ($y_0$). Given a halo proxy $\mathcal{O}$, the mass to observable relation\footnote{Note that although we focus here on outskirt mass $M_{\ast, [50, 100]}$, the same methodology could also be applied to galaxy total mass or any other aperture mass.} is: 
\begin{equation}
\mathrm{log_{10}} (\mathcal{O})=\mathcal{N}\left(\beta~\mathrm{log_{10}}(M_h) + y_0, \sigma_{\mathrm{log_{10}} (\mathcal{O})(M_h)}\right).
\label{eq:lognormal}
\end{equation}
\citetalias{huang2021outer} do not provide the calibrated mass to observable relation for $\m50100$. Here, we begin by deriving an SHMR for $\m50100$ and $\lambda$ that matches the bins in $\m50100$ and richness from \citetalias{huang2021outer} along with their respective underlying halo distributions. In the following subsections, we describe how this goal is achieved. Note that we use the {\it total} scatter in the mass to observable ratio. This includes the intrinsic scatter of the relation convolved with the statistical uncertainties of stellar mass and richness measurements.

\subsubsection{The halo mass to $\m50100$ relation}

\begin{figure*}
    \centering
    \includegraphics[width=\textwidth]{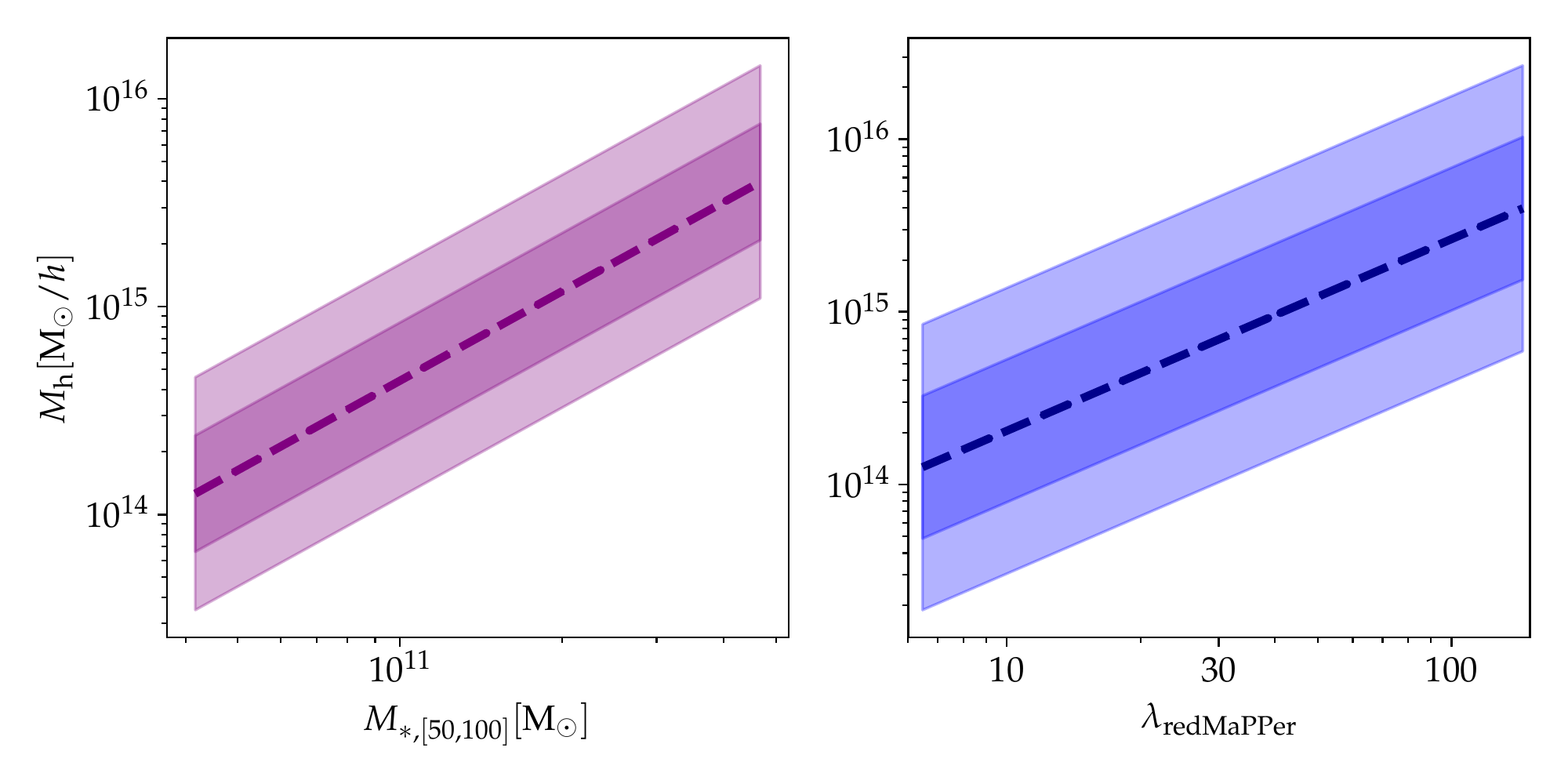}
    \caption{The mass to observable relations for massive galaxies (left) and clusters (right). The dashed lines show the mean relations while the darker and lighter colors show the 1 and 2$\sigma$ spread of the relation. The scatter is a combination of intrinsic and measurement error from the observations of \citetalias{huang2021outer}. The left-hand panel shows the relation for the galaxy outskirt mass.}
    \label{fig:mass_to_obs_relations}
\end{figure*}

From measurements based on the halo mass predictions of the ASAP model, we derive the slope of the halo mass to $\m50100$ relation to be $\beta_{[50, 100]} = 0.7$ \citep{h18, h20, huang2021outer}. We tune the remaining free parameters in Eq.~\eqref{eq:lognormal} to reproduce the halo mass distribution of  \citetalias{huang2021outer}. For consistency we use the same simulation as in \citetalias{huang2021outer}, namely Multi Dark Plank 2 \citep[MDPL2,][]{Prada2011HaloCosmology} at $z = 0.37$. We generate mocks by populating MDPL2 halos with galaxies according to the log-linear relation in Eq. \ref{eq:lognormal}. We calibrate the mass-to-observable relation assuming only central galaxies and utilizing $M_\mathrm{200b}$ as the proxy for halo mass. We fix the slope at $\beta_{[50, 100]} = 0.7$ and vary $\sigma_{\mathrm{log} M_\ast}$ and $y_0$. For each mock, we compute the underlying distribution of halo mass for the same bins in $\m50100$ as in \citetalias{huang2021outer}. We pick values for $\sigma_{\mathrm{log} M_\ast}$ and $y_0$ that best match the halo mass bins from \citetalias{huang2021outer}. For $\m50100$, the best fit parameters are $y_0 = 0.75 $ and $\sigma_{\mathrm{log} M_\ast} = 0.4$, and the calibrated halo mass to $\m50100$ relation is: 
\begin{equation}
\mathrm{log_{10}} \m50100 =\mathcal{N}\left(0.7~\mathrm{log_{10}}(M_h) + 0.75, 0.4\right).
\label{eq:lognormal_m50100}
\end{equation}
The left panel of Figure \ref{fig:mass_to_obs_relations} displays this relation.

\subsubsection{The halo mass to richness relation}

Previous studies have modeled the mass-richness relation with a simple log-linear function \citep[e.g.][]{desi_19} with a slope of $\beta_\lambda = 0.9$. To calibrate the other free parameters ($\sigma_{\mathrm{log_{10}} \lambda}$ and $y_0$), we follow the same methodology as above. Assuming a pivot halo mass of $M_\mathrm{pivot} = 3\times 10^{14} \mathrm{M_\odot}$, the mass to richness relation (Figure \ref{fig:mass_to_obs_relations}, right panel) is: 
\begin{equation}
\mathrm{log_{10}} \lambda =\mathcal{N}\left(0.9~\mathrm{log_{10}}(M_h/M_\mathrm{pivot}) + 1.15, 0.46\right).
\label{eq:lognormal_richness}
\end{equation}

Note that $\sigma_\lambda$ is larger than $\sigma_\mathrm{log \m50100}$ (Equation \ref{eq:lognormal_m50100}). This is the result of the degeneracy between $\beta$ and $\sigma_{\mathrm{log_{10}} (\mathcal{O})}$ \citepalias{huang2021outer}. Different combinations of $\beta$ and $\sigma_{\mathrm{log_{10}} M_\ast}$ can yield similar distributions in halo mass. Because the slopes of the mass to observable relations are different, a relative comparison of $\sigma_{\mathrm{log_{10}}\m50100}$ and $\sigma_{\lambda}$ does not provide any meaningful information. For this project, we calibrate both mass to observable relations to replicate the same halo mass distributions as in \citetalias{huang2021outer}. We illustrate this further in Section \ref{sec:fid_data_vector} and Figure \ref{fig:halo_mass_dist}.

\subsection{The Stellar Mass Function}

The stellar mass function $(\Phi\left(\mathcal{O} \mid M_{h}\right)$, hereafter SMF) describes the number of galaxies (clusters)\footnote{In this section, we use the term galaxies and clusters interchangeably as this derivation is valid for both tracers.} with halo proxy, $\mathcal{O}$, in the range $\mathcal{O} \pm d\mathcal{O}/2$, at constant halo mass. To measure the SMF, we start by computing the {\it conditional} SMF: 
\begin{displaymath}
  \Phi_{c}\left(\mathcal{O} \mid M_{h}\right) =\frac{1}{\ln (10) \sigma_{\log \mathcal{O}} \sqrt{2 \pi}} \times \hspace{0.65\columnwidth}
\end{displaymath}
\begin{equation}
\label{conditional_smf}
\exp \left[-\frac{\left[\log _{10}\left(\mathcal{O}\right)-\log _{10}\left(f_{\mathcal{O}}\left(M_{h}\right)\right)\right]^{2}}{2 \sigma_{\log \mathcal{O}}^{2}}\right].
\end{equation}
The average count of galaxies (clusters) within the bin [$\mathcal{O}^{t_1},\mathcal{O}^{t_2}$] hosted by a halo of mass $M_h$ can be computed through the halo occupation function:
\begin{equation}
\left\langle N_{\mathrm {cen }}\left(M_{h} \mid \mathcal{O}^{t_{1}}, \mathcal{O}^{t_{2}}\right)\right\rangle=\int_{\mathcal{O}^{t_{1}}}^{\mathcal{O}^{t_{2}}} \Phi_{c}\left(\mathcal{O} \mid M_{h}\right) \mathrm{d} \mathcal{O}.
\end{equation}

Finally, the SMF, the abundance of galaxies in a stellar mass bin, can be computed from the halo mass function and the halo occupation function: 
\begin{displaymath}
\Phi_{\rm SMF}(\mathcal{O}^{t_1},\mathcal{O}^{t_2}) \hspace{0.65\columnwidth}
\end{displaymath}
\begin{eqnarray}
&=&\int_{0}^{\infty} \langle N_{\rm cen}(M_h|\mathcal{O}^{t_1},\mathcal{O}^{t_2}) \rangle \frac{{\rm{d}}n}{{\rm{d}}M_h}{\rm{d}}M_h,
\end{eqnarray}
where as a halo mass function, $\frac{{\rm{d}}n}{{\rm{d}}M_h}$, we use \citet{tinker_08}.

\subsection{Power Spectra}

In the halo model, the cross-galaxy--matter power spectrum can be written as the sum of the one and two-halo terms defined as follows:
\begin{align}
P_{\rm gm, 1h}(k,z) &= \frac{ \int dM  \frac{dn}{dM}\frac{M}{\bar{\rho}} \tilde{u}_{\mathrm{m}}(k,M) \langle N_\mathrm{cen} (M_h| \mathcal{O}^{t_1}, \mathcal{O}^{t_2}) \rangle}{\Phi_{\rm SMF}(\mathcal{O}, z)} \hspace{0.65\columnwidth} \notag\\
P_{\rm gm, 2h}(k,z) &= b_{\mathcal{O}}(z) P_{\mathrm{lin}}(k,z). \hspace{0.5\columnwidth}
\label{eq:Pgm}
\end{align}
Here, $ \tilde{u}_{\mathrm{m}}(k, M)$ is the Fourier transform of the radial matter density profile of a halo mass $M_h$. $\tilde{u}_{\mathrm{m}}(k,M)$, is modeled assuming a Navarro, Frenk \& White profile \citep[NFW,][]{nfw} with the \citet{diemer_19} mass-concentration relation $c(M,z)$.

In the two halo term, $P_{\mathrm{lin}}(k,z)$ is the linear matter power spectrum, while $b_{\mathcal{O}}(z)$ is the mean linear bias of galaxies (clusters) computed as: 
\begin{equation}
\label{eq:blambda}
b_{\mathcal{O}}(z) =  \frac{ \int dM  \frac{dn}{dM} b_{\mathrm{h}}(M)\langle N_\mathrm{cen} (M_h| M^{t_1}_*, M^{t_2}_*) \rangle}{{\Phi_{\rm SMF}(\mathcal{O}, z)}}\,,
\end{equation}
where $b_{\mathrm{h}}(M)$ is the halo bias relation in \citet{tinker_10}.

The galaxy--galaxy power spectrum can be modeled as follows: 
\begin{equation}
P_{\rm gg}(k,z) = b_{\mathcal{O}}(z)^2 P_{\mathrm{lin}}(k,z).
\end{equation}

\subsection{Observable Data Vectors}

We derive our three data vectors ($\ds$, $w_p$ and $\nbar$)\footnote{Derivations in this section are adapted from \citet{vandenBosch2013_MNRAS_430_725}.} from the power spectra and stellar mass function introduced previously. $\ds$ and $w_p$ are obtained from the galaxy--matter and galaxy--galaxy two-point correlation functions respectively. These are just the inverse Fourier transform of the galaxy--matter and galaxy--galaxy power spectra: 
\begin{equation}
\xi_{\rm ab}(r,z)={1 \over 2 \pi^2} \int_0^{\infty} P_{\rm ab}(k,z) {\sin kr \over kr} \, k^2 \rmd k\,. 
\label{xiFTfromPK}
\end{equation}
Here, $a$ and $b$ represent the galaxy and matter components of the power spectrum. 

To compute $\ds$, we begin with the surface density, $\Sigma$. The surface density is the integral of the galaxy--matter correlation function along the line of sight: 
\begin{equation}\label{Sigma}
\Sigma(R,z) = \bar{\rho}_\rmm \int_{0}^{\omega_\rms} 
\left[1+\xi_{\rm gm}(r,z)\right] \, \rmd \omega \,.
\end{equation}
$\ds$ is the excess surface density along the line of sight. Therefore it can be computed via:
\begin{equation}\label{shearb} 
\Delta\Sigma (R,z) = \overline{\Sigma}(<R,z) - \Sigma(R,z)\,,
\end{equation}
where $\bar{\Sigma}(<R|z)$ is the mean surface density spanning from $R'=0$ to $R'=R$: 
\begin{equation}\label{averageSigma}
\bar{\Sigma}(<R|z)  = {2\over R^2}
\int_0^R \Sigma(R'|z) \, R' \, \rmd R'\,.
\end{equation}

To compute $w_p$, we integrate the galaxy--galaxy correlation function along the line of sight \citep{davis_peebles_83, vandenBosch2013_MNRAS_430_725}:
\begin{equation}
w_{\mathrm{p}}\left(r_{\mathrm{p}}, z\right)=2 \int_{r_{\mathrm{p}}}^{\infty} \xi_{\mathrm{gg}}(r, z) \frac{r \mathrm{d} r}{\sqrt{r^{2}-r_{\mathrm{p}}^{2}}}.
\end{equation}

Finally, the galaxy (cluster) number density, $\nbar$ is equal to the SMF measured between $\mathcal{O}^{t_1}$ and $\mathcal{O}^{t_2}$ :
\begin{equation}
    \nbar = \Phi_{\rm SMF}(\mathcal{O}^{t_1},\mathcal{O}^{t_2}).
    \label{eq:nbar}
\end{equation}

\section{Method}\label{method}

\subsection{Model Implementation}

Our theoretical framework is implemented in {\cosmolike} \citep{cosmolike}. {\cosmolike} is a code package that aims to constrain cosmology through fast likelihood analysis of joint cosmological probes. In our case, these probes are $\ds, \wpp$, $\bar{n}$, and are derived assuming the halo model and a fiducial cosmology as described in Sec. \ref{theory}. To constrain cosmological parameters, we follow a Bayesian approach when fitting a fiducial data vector given a joint covariance. We describe the joint covariance in Section \ref{errors} and our fiducial sample in Section \ref{sec:fid_data_vector}. 

\subsection{Covariance matrix}\label{errors}

To calculate a covariance matrix, we use the AbacusSummit covariance simulation suite \citep{abacus_1}. These are a suite of dark matter-only simulations with box size $500\, \mathrm{Mpc}/h$ and particle mass $m_\mathrm{p}=2\cdot10^9~\mathrm{M\odot}/h$. The set of covariance boxes includes 2000 different realizations, $15\%$ of which are unavailable at the redshift of interest. Thus, for this project, we use 1700 boxes at $z=0.2$. We generate galaxy and cluster mock catalogs using the {\fontfamily{lmtt}\selectfont CompaSo} halo catalogs available for this suite \citep{Hadzhiyska_21b}. First, we select all host halos. Then, we assign a central galaxy or cluster to host halos using a mass-to-observable relation, including scatter (Section \ref{sec:mass_to_observable}). We create bins in stellar mass and richness, corresponding to our fiducial samples (Table \ref{tab:mass_bins}). Then, we compute our data vectors from all 1700 mocks. 

We build galaxy--galaxy lensing profiles through the excess surface density of dark matter particles in cylinders surrounding host halos. This is implemented in the {\fontfamily{lmtt}\selectfont mean\_delta\_sigma} function of {\fontfamily{lmtt}\selectfont Halotools} \citep{Hearin2017ForwardHalotools}. We compute $\wpp$ by integrating the redshift space two-point correlation function up to a $\pi_\mathrm{max} = 100\mathrm{Mpc}/h$ implemented in the {\fontfamily{lmtt}\selectfont Halotools} function {\fontfamily{lmtt}\selectfont w\_p}.

\begin{figure}
    \centering
    \includegraphics[width=\columnwidth]{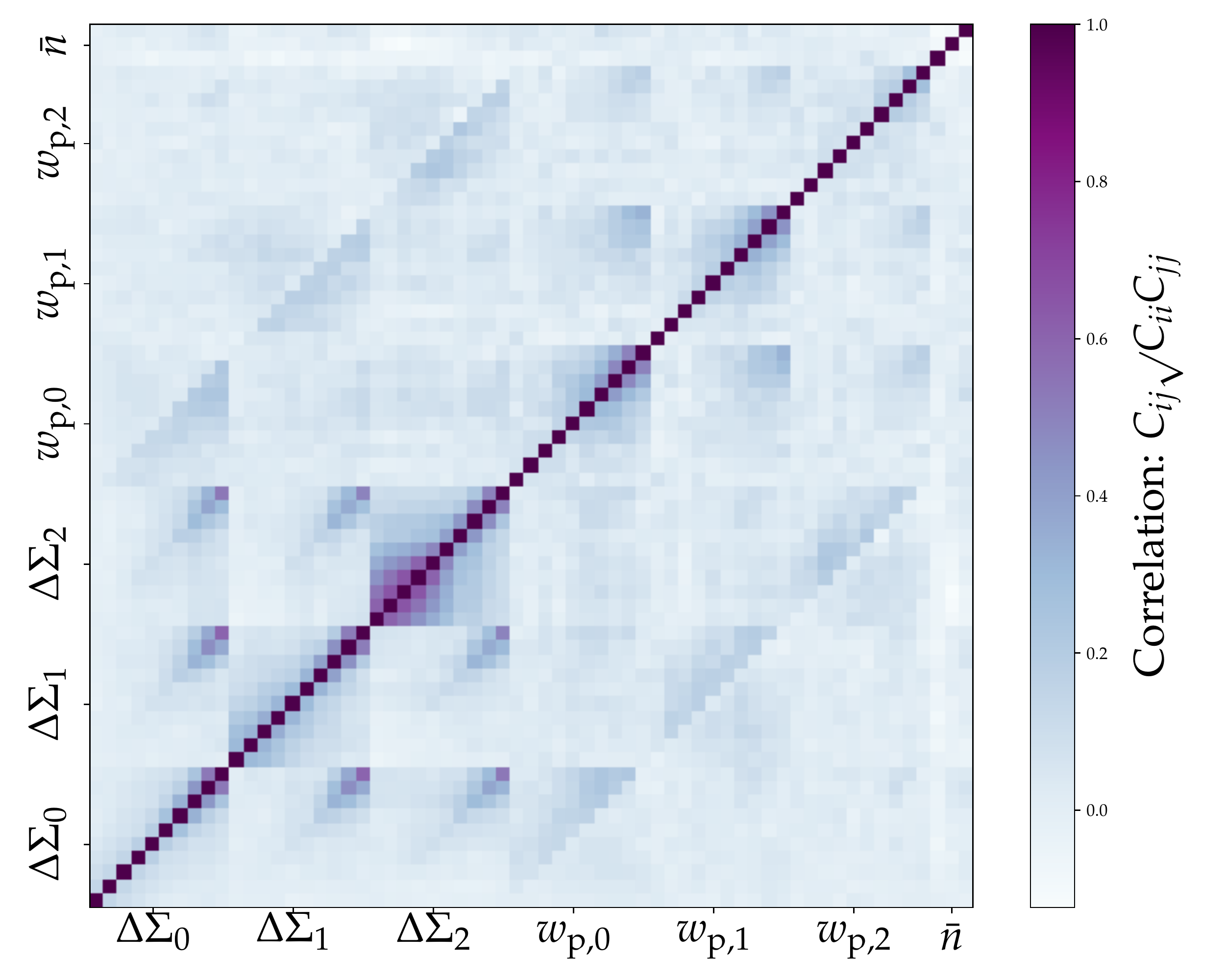}
    \caption{The assumed covariance matrix for the lensing, projected clustering, and number density measurements across three different $M_\ast$ bins for the $\sbin$ samples as measured from the Abacus simulations.}
    \label{fig:corrcof}
\end{figure}

We build a covariance matrix using the mock data vectors from all 1700 mock realizations and show the results in Figure \ref{fig:corrcof}. There are two additional steps to obtain the covariance matrix for our analysis. First, for simplicity, we assume that $\Delta\Sigma$ and $\wpp$ are drawn from the same survey area of 1000 deg$^2$ (although, in practice, for DESI, these areas will be different). We rescale the covariance matrix to the HSC area. The scaling factor is computed following: 
\begin{equation}
    \mathcal{F}_\mathrm{rescale} = \large{(} V_{z2}-V_{z1} \large{)}\frac{A_\mathrm{HSC}}{A_\mathrm{sky}},
\label{eq:rescaling_factor}
\end{equation}
where $V_{z1}$ is the comoving volume to $z=0.3$, $V_{z2}$ is the comoving volume to $z=0.6$, $A_\mathrm{HSC}$ is the HSC survey area (1000 deg$^2$), and $A_\mathrm{sky}$ is the full sky area, $A_\mathrm{sky}=41253$ deg$^2$. Second, we include lensing shape noise in the $\Delta\Sigma$ data vector. We compute the shape noise analytically as in \citet{Singh2017Galaxy-galaxyProperties}. Finally, we add the diagonal terms of the shape noise to the simulation covariance.

\subsection{Fiducial Galaxy and Cluster Samples}\label{sec:fid_data_vector}

We define several samples in a single redshift bin $z \in [0.3, 0.6]$. For most of this paper, we assume samples drawn from the intersection of HSC wide and DESI and assume a survey area of 1000 deg$^2$. The primary goal of this paper is a {\it relative} comparison of the statistical constraining power of different techniques, and so the exact area assumed is not of great importance. The use of the fiducial samples is twofold: We define the fiducial samples in our analytical model used to perform the cosmological analysis; we also define them in the mock realizations from the Abacus suite. The latter is necessary in order to build the covariance. 

\begin{figure*}
    \centering
    \includegraphics[width=\textwidth]{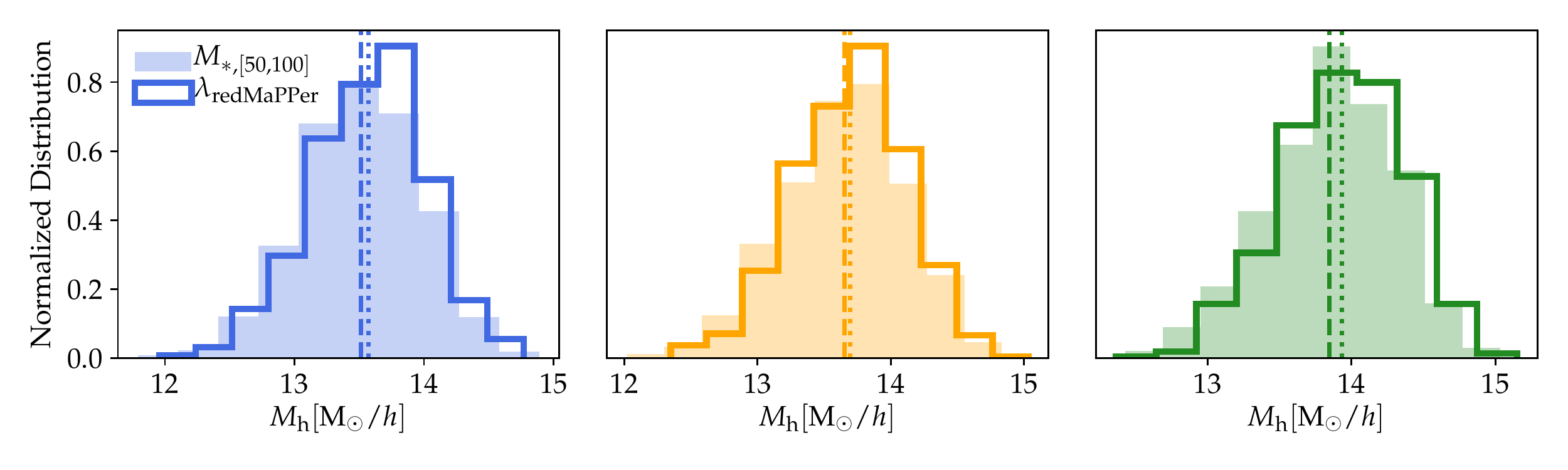}
    \caption{The underlying halo mass distribution for the three fiducial bins in $\m50100$ (filled) and $\lambda$ (step) from Phase 3000 of AbacusSummit simulation mocks at $z=0.2$. Dotted and dashed lines show the mean halo mass of each bin for $\m50100$ and $\lambda$, respectively. The underlying halo mass distribution for massive galaxies and clusters is similar across all three bins.}
    \label{fig:halo_mass_dist}
\end{figure*}

We consider three sets of samples. The first set corresponds to the super-massive galaxy and cluster samples used in \citetalias{huang2021outer}, hereafter $\sbin$. This set consists of 3 samples binned by outskirt mass and $\lambda$ (see Table \ref{tab:mass_bins} for the bins' properties). We illustrate the underlying halo mass distribution of $\sbin$ in Figure \ref{fig:halo_mass_dist}. These distributions are built with a mock realization of the AbacusSummit (specifically Phase 3000 at $z=0.2$; see Section \ref{errors} for the description of our mocks). The $\m50100$ and $\lambda$ bins in Figure \ref{fig:halo_mass_dist} have similar underlying mass distributions. This is by design and is motivated by the work of \citetalias{huang2021outer}. As shown in \citetalias{huang2021outer}, galaxy outskirt mass is an excellent proxy of halo mass. Hence, it is possible to construct samples binned by outskirt mass with similar underlying halo mass distributions to ones binned by $\lambda$. Based on early analysis of DESI data, we expect that DESI will obtain complete samples of galaxies in the mass range probed by these bins.

$\sbin$ extends to a lower halo mass than the fiducial sample of the DES cluster cosmology analysis \citep{desi20_clusters}. The DES cluster sample ranges from $\lambda=20$ to $\lambda=200$. This corresponds to our highest richness bin. The lowest limit on our cluster sample is $\lambda=9.22$. We believe pushing to lower halo mass samples may be possible with DESI, as the DES sample was limited by cluster-finding systematics. In particular, these affect the low richness regime the most and can be notoriously difficult to accurately model and marginalize over \citep{DeRose2019, Korytov2019, sunayama_2020S, joe_22, wu_22, park_23}. In contrast, pushing to lower halo masses with the outskirt mass tracer may be more straightforward, although future work is needed to determine if this is indeed possible. Pushing to lower halo masses (higher number densities) will improve the signal-to-noise of our data vector (especially $\wpp$), leading to tighter cosmological constraints.

Second, we consider a galaxy sample that combines all galaxies of $\sbin$ into a single bin. We name this sample $\sfull$. This will be a cumulative threshold sample that will include all super-massive galaxies with stellar mass above $M_\ast = 11.5 \mathrm{M}_\odot$. Because DESI will be complete above $M_\ast = 11.5 \mathrm{M}_\odot$, $\sfull$ will also be a mass complete sample. This sample aims to study the impact of binning on our cosmological constraints. 

Finally, we also assume a sample of galaxies with the same number density as the DESI Bright Galaxy Survey (BGS) sample at $z \sim 0.3$, hereafter $\sbgs$. BGS is a magnitude-limited sample ($r < 19.5$) at low redshift with $\nbar = 1.70 \times 10^{-3}~h^3/\mathrm{Mpc}^3$ \citep[][]{bgs_desi}. Therefore $\sbgs$ has a much higher number density than $\sfull$ ($\nbar = 2.36 \times 10^{-5}~h^3/\mathrm{Mpc}^3$). However, the full BGS sample will not be mass-complete. We do not expect to measure outskirt masses for the full BGS sample for two reasons. [1] To measure the outskirt mass, we use a fixed physical kpc boundary. This means we cannot apply it to the high number density regime since 50 kpc is too large for low-mass galaxies. [2] Extending our sample to low-mass galaxies would mean that we need to introduce a break in our SHMR model, and we are not currently doing this. The purpose of our BGS-like sample is solely to compare the constraining power of our massive galaxies sample ($\sbin$) with a larger but mass incomplete sample ($\sbgs$). Thus, in the context of this work, the BGS-like sample is a comparison sample that aims to understand the impact of mass completeness on cosmological constraints. All three samples and their properties are displayed in Table \ref{tab:mass_bins}. 

\begin{figure}
    \centering
    \includegraphics[width=\columnwidth]{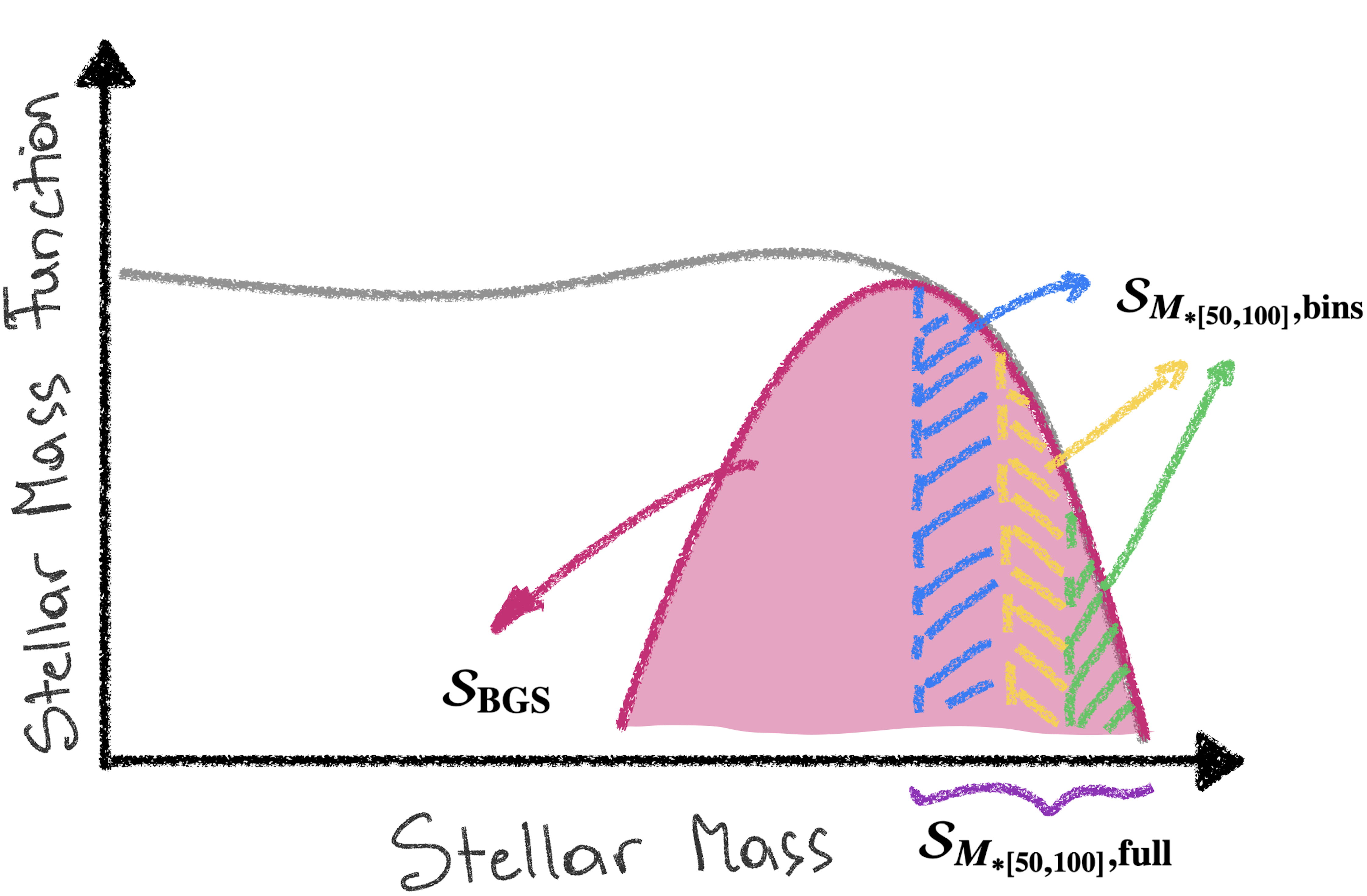}
    \caption{Illustration of the stellar mass function for our three sets of galaxies. The $\sbin$ samples (blue, orange, green) are binned by galaxy mass in a range where DESI will be complete in stellar mass. In this mass range, we can use number densities for constraints, similar to how cluster abundances are used in cluster cosmology. The second sample, $\sfull$, is one threshold-limited sample that combines all three bins from $\sbin$. The third sample is $\sbgs$ (pink). This sample has a higher number density and extends to lower halo masses but is incomplete in galaxy mass.}
    \label{fig:smf_illustration}
\end{figure}

Figure \ref{fig:smf_illustration} illustrates the different samples. The $\sbin$ is designed to study how samples of galaxies binned by outskirt mass compared with samples binned by $\lambda$. The $\sfull$ sample is designed to study the impact of binning. Finally, $\sbgs$ allows us to study the trade-off between samples that are large but incomplete and for which number density cannot be used in the cosmological parameter fitting versus samples that have overall lower number densities but which are complete, and for which number densities can be used for cosmological parameter fitting (similar to cluster cosmology analyses). Throughout this paper, the $\sbin$ is our fiducial sample, and we study how cosmological constraints from this sample compare with constraints from the other samples.

\setlength{\tabcolsep}{20pt}
\renewcommand{\arraystretch}{1.3}

\begin{table*}
\caption{Fiducial richness and outer stellar mass bins for low, medium, and high halo masses (top row), a cumulative threshold sample (mid rows), and a DESI BGS-like sample (bottom row) along with the respective number densities. Note that the stellar mass values shown here are measured in outskirt mass rather than the total stellar mass. A galaxy with $\m50100 \approx 10^{10.85} \mathrm{M}_\odot$ resides in a halo with mass $M_h \approx 10^{13.5} \mathrm{M}_\odot/h$.}
\begin{tabular}{ccccc}
\toprule
\multicolumn{1}{|c|}{\begin{tabular}[c]{@{}c@{}}$\boldsymbol{\mathrm{log}_{10}M_{\ast [50, 100]}}$ \\ $\boldsymbol{[\mathrm{log}_{10} \mathrm{M}_\odot/h]}$\end{tabular}} &
  \multicolumn{1}{c|}{\begin{tabular}[c]{@{}c@{}}$\boldsymbol{\mathrm{log}_{10}M_h}$ \\ $\boldsymbol{[\mathrm{log}_{10} \mathrm{M}_\odot/h]}$\end{tabular}} &
  \multicolumn{1}{c|}{\begin{tabular}[c]{@{}c@{}}$\boldsymbol{n_\mathrm{gal}}$\\ $\boldsymbol{(\mathrm{Mpc}/h)^{-3}}$\end{tabular}} &
  \multicolumn{1}{c|}{\begin{tabular}[c]{@{}c@{}}$\boldsymbol{n_\mathrm{gal, cummulative}}$\\ $\boldsymbol{(\mathrm{Mpc}/h)^{-3}}$\end{tabular}} &
  \multicolumn{1}{c|}{$\boldsymbol{\lambda}$ bins} \\ \hline
\multicolumn{5}{|c|}{$\large\boldsymbol{\mathcal{S}_{M_{\ast [50, 100]}, \mathrm{bins}}}$} \\ \hline
\multicolumn{1}{|c|}{{[}10.8, 10.95)} &
  \multicolumn{1}{c|}{$13.51 ^{+0.47}_{-0.47}$} &
  \multicolumn{1}{c|}{$1.09\cdot 10^{-5}$} &
  \multicolumn{1}{c|}{$1.09\cdot 10^{-5}$} &
  \multicolumn{1}{c|}{{[}9.22-13.5)} \\
\multicolumn{1}{|c|}{{[}10.95, 11.10)} &
  \multicolumn{1}{c|}{$13.69^{+0.51} _{-0.44}$} &
  \multicolumn{1}{c|}{$6.83\cdot 10^{-6}$} &
  \multicolumn{1}{c|}{$1.77\cdot 10^{-5}$} &
  \multicolumn{1}{c|}{{[}13.5-21.00)} \\
\multicolumn{1}{|c|}{{[}11.10, 12.10{]}} &
  \multicolumn{1}{c|}{$13.87^{+0.49}_{-0.43}$} &
  \multicolumn{1}{c|}{$5.95\cdot 10^{-6}$} &
  \multicolumn{1}{c|}{$2.36\cdot 10^{-5}$} &
  \multicolumn{1}{c|}{{[}21.00-150{]}} \\ \hline
\multicolumn{5}{|c|}{$\large\boldsymbol{\mathcal{S}_{M_{\ast [50, 100]}, \mathrm{full}}}$} \\ \hline
\multicolumn{1}{|c|}{{[}10.8, 12.1{]}} &
  \multicolumn{1}{c|}{$13.65^{+0.50}_{-0.47}$} &
  \multicolumn{1}{c|}{$2.36\cdot 10^{-5}$} &
  \multicolumn{1}{c|}{$2.36\cdot 10^{-5}$} &
  \multicolumn{1}{c|}{{[}9.22, 150{]}} \\ \hline
\multicolumn{5}{|c|}{$\large\boldsymbol{\mathcal{S}_\mathrm{BGS}}$} \\ \hline
\multicolumn{1}{|c|}{{[}9.97, 12.1{]}} &
  \multicolumn{1}{c|}{$12.78^{+0.60}_{-0.61}$} &
  \multicolumn{1}{c|}{$1.70 \cdot 10^{-3}$} &
  \multicolumn{1}{c|}{$1.70 \cdot 10^{-3}$} &
  \multicolumn{1}{c|}{-} \\ \bottomrule
\end{tabular}
\label{tab:mass_bins}
\end{table*}

\subsection{Radial range of data vectors}\label{range}

 We use {\cosmolike} to predict the projected observables, $\ds$ and $\wpp$, in 10 logarithmically distributed radial bins ($R_\mathrm{p}$) ranging from 0.4 to 40 $\mathrm{Mpc}/h$. The number of radial bins is determined such that the Hartlap factor is minimal ($\sim 3\%$) \citep{Hartlap:2007}. Since we use data vectors of different lengths throughout the paper, the Hartlap factor will change. However, we expect this change to be at the percent level due to the large number of realizations we are using (1700 different phases of the Abacus Suite). Both the upper and lower radial limits are determined by the AbacusSummit Simulations \citep{Maksimova_2021}, which we used to compute the Gaussian and non-Gaussian components of the covariance (Section \ref{errors}). The lower limit is restricted by the resolution of the simulations, while the upper limit is set by the box size (500 Mpc/$h$).
 
\begin{table}
\centering
\caption{Model parameters and priors used in the joint likelihood analysis introduced in Section \ref{s:mcmc_fitting}. The top row describes the cosmological parameters, while the two bottom rows describe the free parameters of the mass to observable relations (Section \ref{sec:mass_to_observable}).}
\begin{tabular}{@{\hskip 0.3in}l@{\hskip 0.7in}c@{\hskip 0.7in}c@{\hskip 0.3in}}
\hline
\textbf{Parameter}           & \textbf{Fiducial} & \textbf{Prior} \\ \hline
$\Omega_m$                   & 0.3      & {[}0.2, 0.9{]} \\
$\sigma_8$                   & 0.8222     & {[}0.6, 1.1{]} \\
$h_0$                        & 0.7      & ---            \\
$\Omega_b$                   & 0.0492   & ---            \\
$n_s$                        & 0.9645   & ---            \\
$\Omega_{\nu, h^2}$          & 0.6844   & ---            \\
$\theta_s$                   & 1.05     & ---            \\
$w_0$                        & -1.0     & ---            \\
$w_a$                        & 0.0      & ---            \\ \hline
$y_{\mathrm{log_{10}} \m50100}$   & 0.75     & {[}0, 2{]}   \\
$\beta_{\mathrm{log_{10}}\m50100}$     & 0.7   & {[}0.5, 1{]}   \\
$\sigma_{\mathrm{log_{10}} \m50100}$ & 0.4    & {[}0.01, 1{]} \\ \hline
$y_{\mathrm{log_{10}} \lambda}$   & 1.15     & {[}0, 2{]}   \\
$\beta_{\mathrm{log_{10}} \lambda}$     & 0.9   & {[}0.5, 1{]}   \\
$\sigma_{\mathrm{log_{10}} \lambda}$ & 0.46    & {[}0.01, 1{]} \\ \hline
\end{tabular}
\label{tab:priors}
\end{table}

\subsection{Model parameters}

\begin{figure*}
    \centering
    \includegraphics[width=\textwidth]{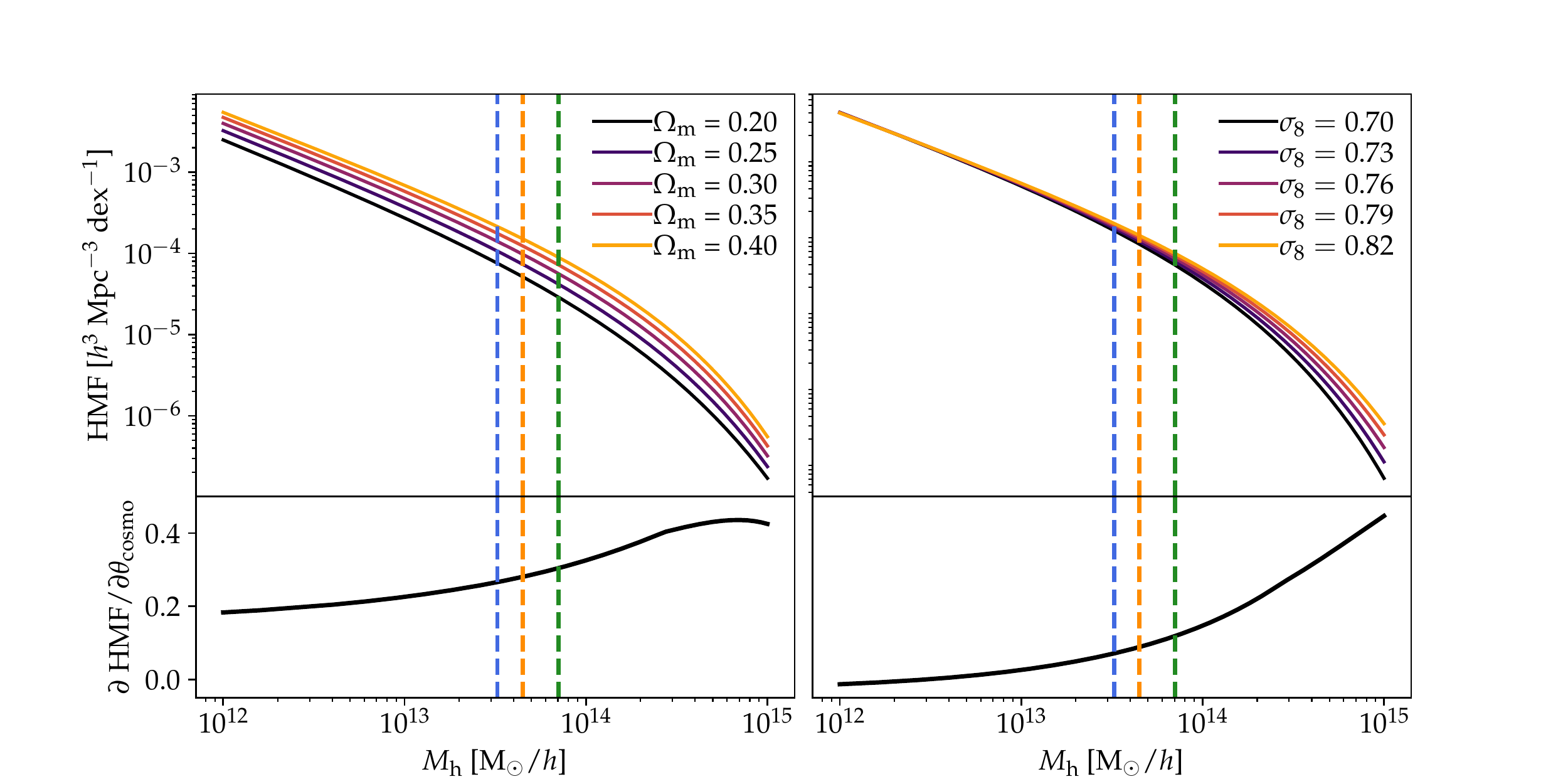}
    \caption{Sensitivity of the HMF to cosmological parameters. Each panel shows how the HMF changes when we vary $\omm$ (left) and $\s8$ (right). $\omm$ changes the overall amplitude of HMF, while $\s8$ affects only the higher mass end. The blue, orange, and green vertical lines indicate the mean values of the stellar mass bins from $\sbin$. The HMF is sensitive to both $\s8$ and $\omm$ in the halo mass range of interest.} 
    \label{fig:change_omm_s8}
\end{figure*}

 Model parameters are divided into two categories. The first set consists of cosmological parameters. We vary $\omm$ and $\s8$ as they are the main parameters influencing the amplitude or the shape of the halo mass function. Other cosmological parameters are fixed to the values in Table \ref{tab:priors}. Figure \ref{fig:change_omm_s8} shows how the halo mass function changes when we vary $\omm$ and $\s8$. $\omm$ affects the overall amplitude, whereas $\s8$ affects the massive end of the halo mass function. At the halo mass range of interest, we should, in principle, be able to constrain both $\omm$ and $\s8$.

\begin{figure*}
    \includegraphics[width=\textwidth]{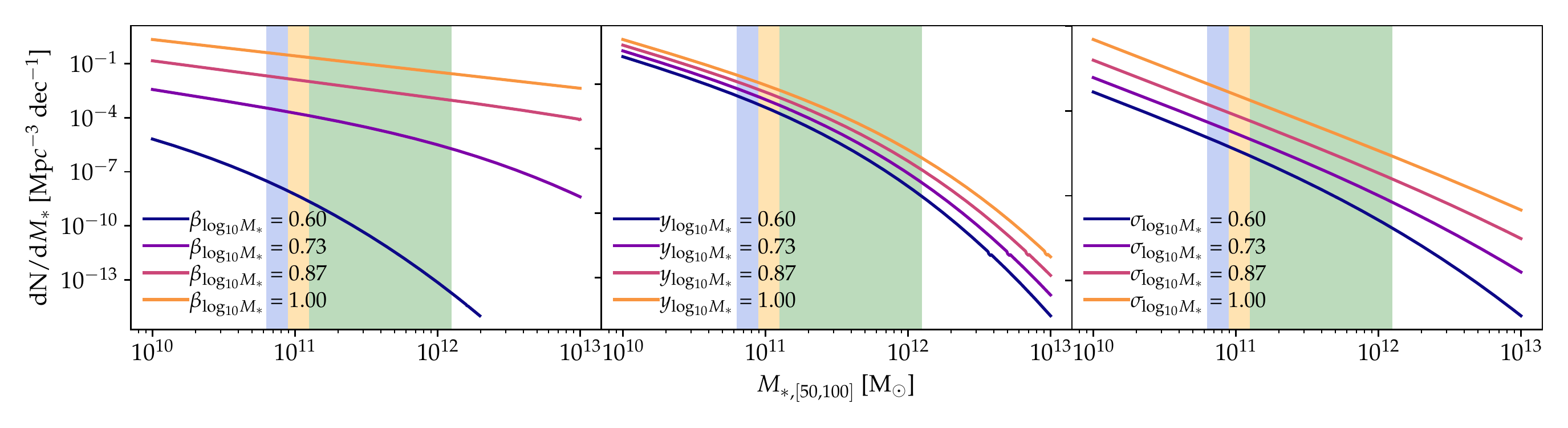}
    \caption{Sensitivity of the SMF to the SHMR parameters. Each panel displays how the SMF changes while varying each of the SHMR parameters while keeping the rest fixed. The shaded regions indicate the stellar mass bins we are interested in (Figure \ref{fig:halo_mass_dist}). All three nuisance parameters affect the amplitude and the shape of SMF.}
    \label{fig:smf_change}
\end{figure*}

The second category of model parameters describes the mass-to-observable relation. These are the slope, the y-intercept, and the scatter in the mass to observable relation. In Figure \ref{fig:smf_change}, we study how the SMF changes as we vary each parameter. All three parameters affect the shape and amplitude of SMF at the stellar mass range of interest. Therefore, we vary all these parameters in our fits.

\subsection{Model Fitting}\label{s:mcmc_fitting}

We follow a Bayesian approach to fit our fiducial data vectors. We sample posteriors from the parameter space following a Markov Chain Monte Carlo (MCMC) approach with {\fontfamily{lmtt}\selectfont emcee } \citep{Foreman-Mackey2012Emcee:Hammer}. We follow the same likelihood analysis as in \citet{cosmolike}, in which the likelihood of the cosmological and nuisance parameters is parametrized jointly as a multivariate Gaussian
\begin{equation}
\mathcal{L}\left(\mathbf{D} \mid \mathbf{p}_{c}, \mathbf{p}_{\mathrm{n}}\right)= \exp \bigg(-\frac{1}{2} \underbrace{\left[(\mathbf{D}-\mathbf{M}\left(\mathbf{p}_{\mathbf{c}}, \mathbf{p}_{\mathbf{n}}\right))^{t} \mathbf{C}^{-1}(\mathbf{D}-\mathbf{M}\left(\mathbf{p}_{\mathbf{c}}, \mathbf{p}_{\mathbf{n}}\right))\right]}_{\chi^{2}\left(\mathbf{p}_{c} , \mathbf{p}_{n}\right)}\bigg).
\end{equation}
Here, {\bf D} is the multi-probe data vector computed at the fiducial cosmological ($\mathbf{p}_c$) and nuisance ($\mathbf{p}_n$) parameters chosen for this paper. {\bf C} is the covariance matrix measured with the Abacus simulations (Figure \ref{fig:corrcof}). Finally, the model {\bf $\mathbf{M}\left(\mathbf{p}_{\mathbf{c}}, \mathbf{p}_{\mathbf{n}}\right)$} is a function of the free parameters.

We use broad, uninformative top hat priors for all free parameters (Table \ref{tab:priors}). We assess convergence by computing the mean auto-correlation time across all dimensions for each MCMC step. We assume that convergence is achieved when the change of the mean auto-correlation time between steps is smaller than 1\% after at least 100 correlation times \citep{Sokal1996MonteCM}\footnote{\url{https://emcee.readthedocs.io/en/stable/tutorials/autocorr/}}.

\section{Results}\label{sec:results}

This paper introduces massive galaxies as competitive halo tracers for future cosmological analyses. Our data vector consists of $\ds, \wpp, \mathrm{and}~\bar{n}$. We assume 1000 deg$^2$ of HSC lensing data with spectroscopic redshifts from DESI and a set of fiducial parameters given in Table \ref{tab:priors}. This section presents our main results. First, we study the impact of binning on cosmological constraints. We then study the impact of $\nbar$ and $\wpp$ on $\s8$ and $\omm$. We then compare how the massive galaxy methodology compares with two other traditional methods: 1) cluster cosmology and 2) joint lensing plus clustering analyses. Finally, we investigate one potential systematic associated with this methodology: the impact of satellites.

\subsection{Massive Galaxies}

\subsubsection{Impact of binning}

We start by analyzing the impact of binning on our constraints. We consider two scenarios: one in which the super-massive galaxies are divided into three narrow bins by outskirt mass ($\sbin$) and another in which all galaxies reside in a single cumulative bin ($\sfull$; see Table \ref{tab:mass_bins} for details on each sample). Both scenarios assume the same survey area (HSC 1000 deg$^2$), fiducial parameters (Table \ref{tab:priors}), and data vectors ($\ds, \wpp, {\rm and}, \nbar$). 

\begin{figure}
    \centering
    \includegraphics[width=\columnwidth]{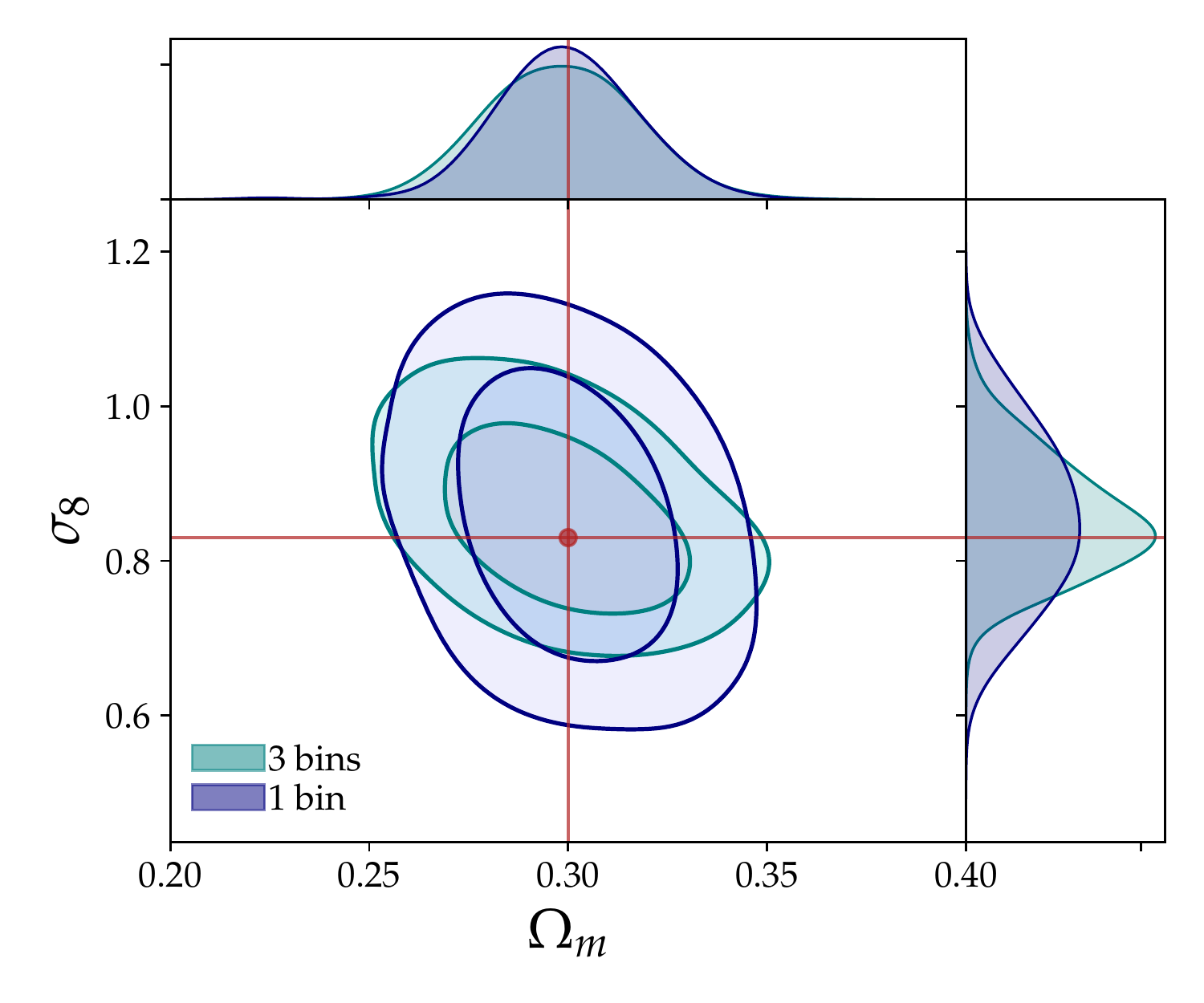}
    \caption{Constraints for $\omm$ and $\s8$ with massive galaxies when considering a single cumulative bin (navy) and three narrow bins (teal). Both analyses assume the same survey area (HSC 1000 deg$^2$) and the same fiducial data vector (Table \ref{tab:priors}). The maroon dot and lines here and throughout the paper show the fiducial values of $\omm$ and $\s8$. Three narrow bins yield 34\% tighter constraints on $\s8$.}
    \label{fig:corner_1_v_3}
\end{figure}

Figure \ref{fig:corner_1_v_3} shows how constraints on $\omm$ and $\s8$ compare for $\sbin$ and $\sfull$. We find that binning by stellar mass does not improve constraints on $\omm$ but yields stronger constraints for $\s8$. The $\sbin$ sample yields $\s8 = 0.86 ^{+ 0.08} _{- 0.06}$ while $\sfull$ yields $\s8 = 0.85 ^{+ 0.12} _{- 0.11}$. Binning by galaxy mass, thus, yields a 34\% improvement over a threshold sample. We can understand this result by looking at the impact of $\s8$ and $\omm$ on the HMF in Figure \ref{fig:halo_mass_dist}. $\omm$ and $\s8$ have different impacts on the HMF: $\omm$ mainly influences the overall amplitude, while $\s8$ induces mass-dependent changes. At $ M_\mathrm{h} \in [10^{13.2}, 10^{14}] \mathrm{M_\odot}\h$, the slope of $\frac{\partial{\rm HMF}}{\partial{\s8}}$ is steeper than the slope of $\frac{\partial{\rm HMF}}{\partial{\omm}}$. Binned samples will better constrain the slope of $\frac{\partial{\rm HMF}}{\partial{\s8}}$. This translates into tighter constraints in $\s8$. 

Our result qualitatively agrees with \citet{wu_12}, who demonstrate that when the bins are correlated with a mass to observable relation, binning does indeed improve constraints in $\s8$. One important assumption here, however, is that the mass-to-observable relation follows a simple log-linear relation with constant scatter and satellite fraction throughout the three bins. While it needs to be studied in greater detail, this is a reasonable assumption for our work as the massive galaxy sample that we adopt covers a very narrow mass range \citep[e.g.][]{desi_19} and well above the pivot mass scales at which the SHMR bends \citep[e.g.,][]{leauthaud_2012}.

\subsubsection{Impact of $\nbar$}\label{sec:nbar_impact}

\begin{figure}
    \centering
    \includegraphics[width=\columnwidth]{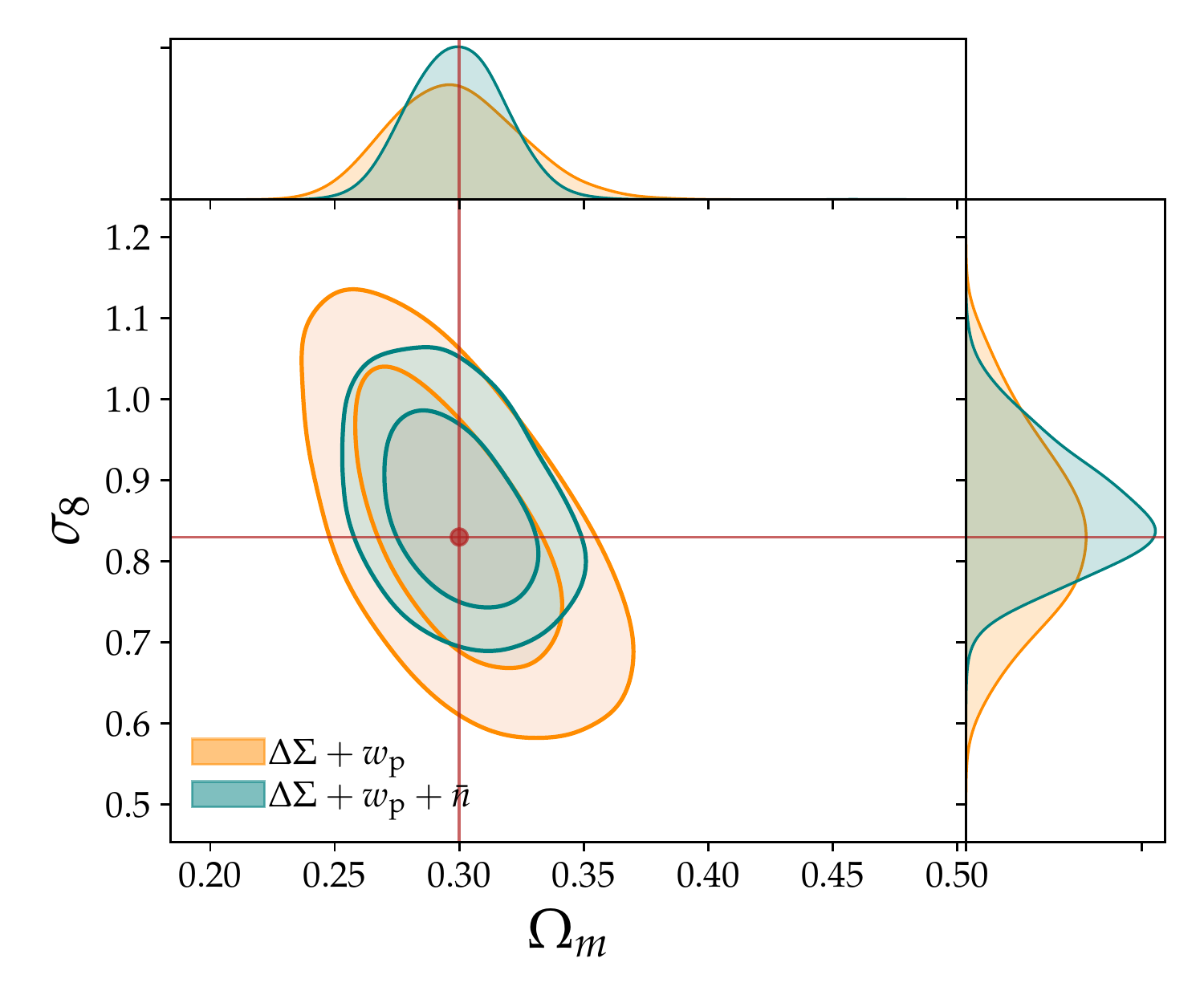}
    \caption{Constraints for $\omm$ and $\s8$ with massive galaxies when including (teal) or omitting (orange) $\nbar$. Both analyses assume the same sample ($\sbin$), survey area (HSC 1000 deg$^2$), and fiducial parameters (Table \ref{tab:priors}). Including $\nbar$ improves constraints of $\s8$ by 33\% and $\omm$ by 23\%.}
    \label{fig:corner_nbar}
\end{figure}

Next, we study the impact of number density on cosmological constraints. Utilizing the sample $\sbin$, we consider two scenarios: one in which we include $\nbar$ in our data vector and another in which we exclude it. The results are displayed in Figure \ref{fig:corner_nbar} (purple contours versus teal contours). Including number density improves $\s8$ constraints by 33\% and $\omm$ by 23\%. It is clear that $\nbar$ affects both cosmological parameters. This has a simple explanation. By definition, $\nbar$ drives the amplitude of the SMF and, thus, the amplitude of the HMF. As both parameters impact the amplitude of the HMF, $\nbar$ helps to constrain both.

\subsubsection{Impact of $\wpp$}

\begin{figure}
    \centering
    \includegraphics[width=\columnwidth]{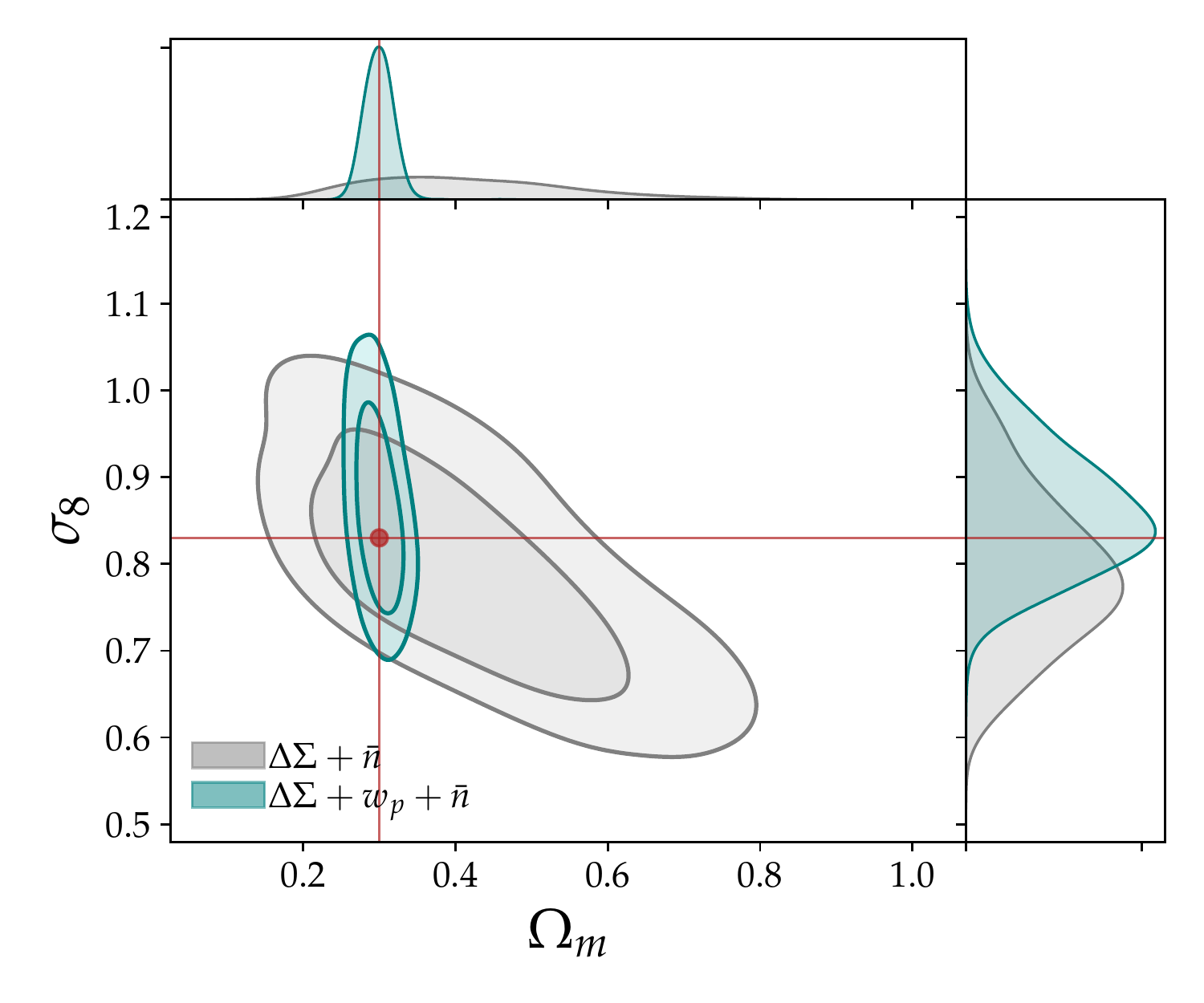}
    \caption{Constraints for $\omm$ and $\s8$ with massive galaxies when including (teal) and omitting (gray) $\wpp$. Both analyses assume the same survey area (HSC 1000 deg$^2$), fiducial parameters (Table \ref{tab:priors}), and the same binning system (Figure \ref{fig:halo_mass_dist}). This assumes a galaxy sample that selects halos with a similar number density as a $\lambda>9.22$ selection. Including $\wpp$ greatly improves constraints on $\omm$, with an improvement of 84\%. This figure presents one of the key potential advantages of massive galaxies over clusters: the ability to push to lower halo masses where $\wpp$ is better-measured.}
    \label{fig:corner_wp}
\end{figure}

Here we analyze the impact of $\wpp$ on cosmological constraints. We adopt $\sbin$ as our baseline sample and consider two cases: one in which we include $\wpp$ as part of the data vector along with $\ds$ and $\nbar$ and another in which we exclude it. Figure \ref{fig:corner_wp} displays the results. Clearly, $\wpp$ substantially impacts $\omm$, improving the constraints by as much as 84\%. This is because the combination of lensing and clustering is historically known to be a powerful cosmology tool, and using the two in combination, helps to break the degeneracy with galaxy bias \citep{yoo_06, Baldauf_2010, more_2013, Mandelbaum2013, Cacciato2013_MNRAS_430_767, more_15, Leauthaud2017_MNRAS_467_3024, Lange2019_MNRAS_488_5771, singh2020}. 

Traditionally, cluster cosmology with optically selected clusters only uses lensing and number densities as their main observables \citep{desi20_clusters, Costanzi_21}. When samples are limited to high-mass halos, the measurement of the clustering of clusters can be noisy and, therefore, may not lead to improvements in cosmological parameter constraints. Only a handful of analyses have explored the clustering of clusters as part of the data vector \citep{mana_13,baxter_16, park_21, to_21}, but in these analyses, additional data are required to constrain cosmological parameters. For example, \citet{to_21} carried out a 4x2 analysis, including the autocorrelation of galaxies and clusters, the cross-correlation of galaxies and clusters, cluster lensing, and cluster abundances. When comparing their constraints with the traditional cluster cosmology results from \citet{Costanzi_21}, they report a 38\% improvement in constraints of $\omm$ and mild improvements in constraints on $\s8$. Qualitatively this agrees with our results, although this improvement is less significant quantitatively. This is mainly due to the different halo mass cuts employed. Indeed, \citet{to_21} use a cluster sample with a richness cut of $\lambda \in [20, 235]$. Instead, our cuts are defined by our estimates for the galaxy mass completeness limit for DESI, which corresponds to $\lambda \in [9.22, 150]$. \textit{This points to a significant advantage in using super massive galaxies over traditional clusters. Namely, super-massive galaxies may enable us to push down in halo mass into the group regime, where the number densities are higher and $\wpp$ offers larger constraining power}.

\subsection{Massive Galaxies vs. Clusters}

Here we compare constraints from clusters versus super-massive galaxies. We assume the same data vector ($\ds, \wpp, \mathrm{and}~\bar{n}$) and adopt three bins of richness and outskirt mass (Figure \ref{fig:halo_mass_dist}). We use the covariance predicted for a 1000 deg$^2$ survey (Section \ref{errors}). 

\begin{figure}
    \centering
    \includegraphics[width=\columnwidth]{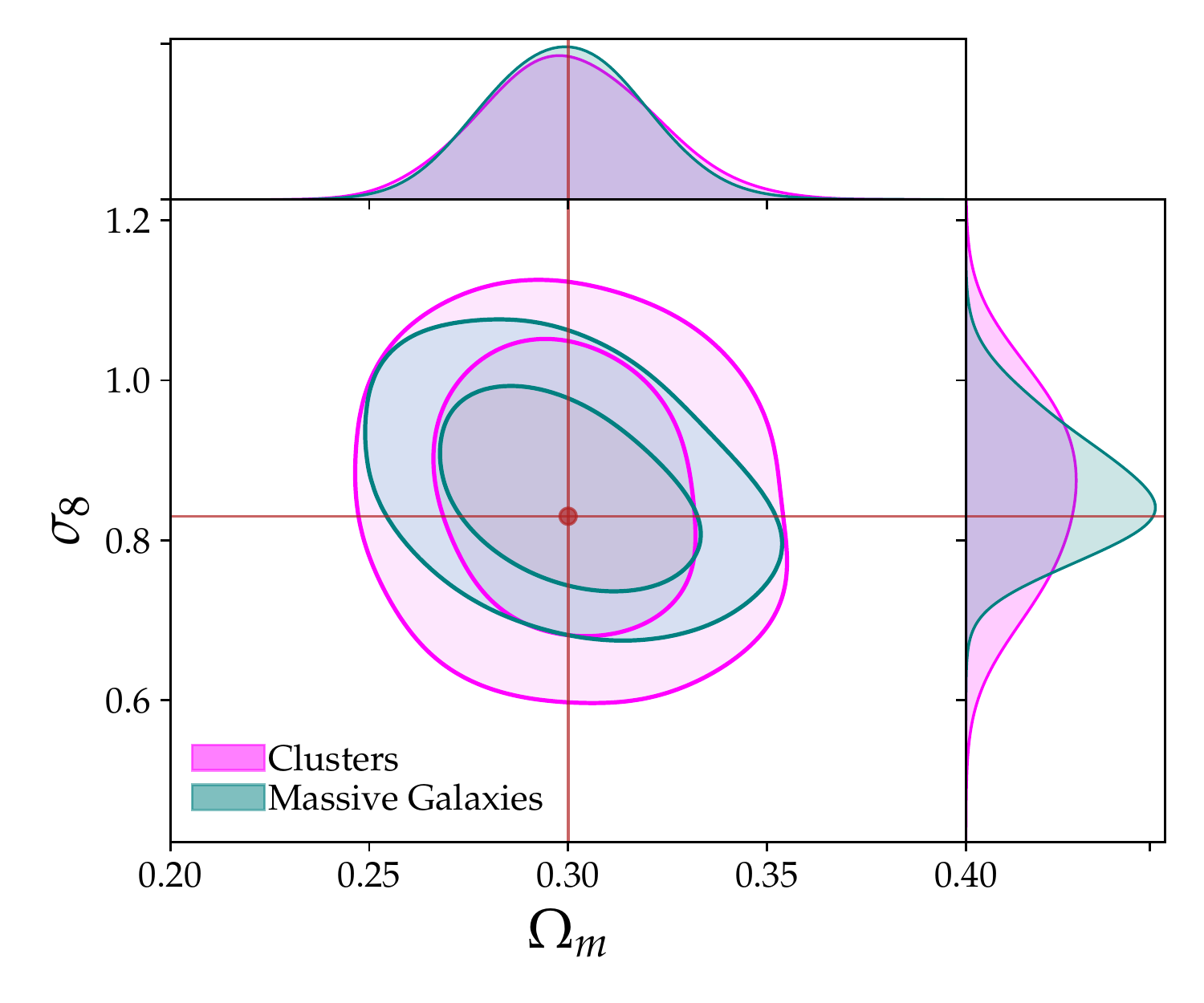}
    \caption{Constraints for $\omm$ and $\s8$ with massive galaxies (teal) and clusters (pink) as halo tracers.
    Both analyses assume the same survey area (HSC 1000 deg$^2$) and fiducial data vector (Table \ref{tab:priors}). Massive galaxies yield competitive constraints while potentially avoiding systematics such as projection effects and miscentering.}
    \label{fig:corner_mg_cl}
\end{figure}

Figure \ref{fig:corner_mg_cl} compares traditional cluster cosmology and massive galaxy cosmology. While constraints in $\omm$ are similar, massive galaxies yield tighter constraints in $\s8$ by 36\%.

The fact that both methods offer similar statistical constraining power can be explained by Figure \ref{fig:halo_mass_dist}. Although the mass-to-observable relations for massive galaxies and clusters are different, they present similar values for the scatter in halo mass at fixed observable (Figure 8, \citetalias{huang2021outer}) and can therefore be used to select similar bins in halo mass. Because they trace the halo mass function in a similar fashion, they also derive comparable constraints for $\s8$ and $\omm$. 

We investigate why massive galaxies yield tighter constraints in $\s8$. The only difference between the cluster and massive galaxies analyses in Figure \ref{fig:corner_mg_cl} is the mass to observable relation. To test which parameters of the mass to observable relation shift the contours, we vary each of them in their respective prior range. We recompute the cosmological constraints for each variation. Our tests conclude that the shift in the contours observed in Figure \ref{fig:halo_mass_dist} is related to differences in the power-law index of the mass-observable relation.

We conclude that massive galaxies can be used as alternatives to clusters to perform similar cosmological analyses. However, massive galaxies may offer advantages over traditional cluster cosmology. These arguments will be presented further in Section \ref{dis:clusters}.

\subsection{The Trade-off Between Complete versus Incomplete Samples}\label{sec:lrgs}

In lensing and clustering cosmological analyses with spectroscopic samples, number density is seldom used as a constraint \citep[e.g.][]{Uitert_18, Joudaki_2018, DESY1, desi_19, singh2020, heymans_21, Miyatake2021_arXiv_2111_2419, desi22, amon22, lange_2023}. This is because galaxy samples from surveys such as BOSS are incomplete, except in specific mass and redshift ranges \citep{leauthaud_16}. In contrast to BOSS, however, DESI will be complete above $\log_{10} M_\ast=11.5~\mathrm{M_\odot}$ over a wide range in redshift which will allow number density to be used as a constraint in a similar fashion to galaxy clusters.

\begin{figure}
    \centering
    \includegraphics[width=\columnwidth]{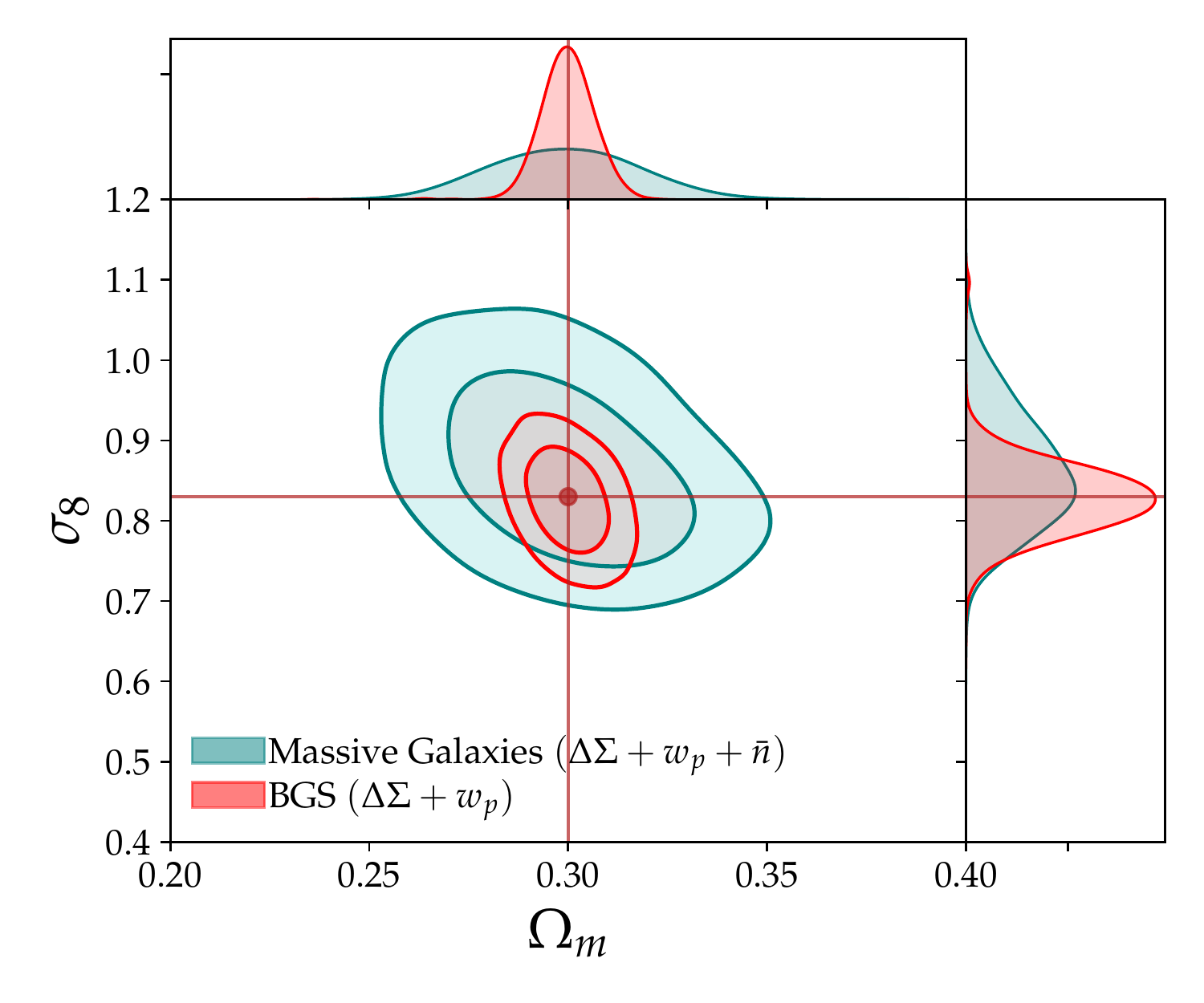}
    \caption{Constraints for $\omm$ and $\s8$ with massive galaxies when $\sbin$ (teal) and $\sbgs$ (red). The $\sbin$ analysis includes all three $\ds, \wpp,~\mathrm{and}~\nbar$ as data vectors, while the $\sbgs$ includes only $\ds~\mathrm{and}~\wpp$. Both analyses assume survey area (HSC 1000 deg$^2$) and fiducial parameters (Table \ref{tab:priors}). The $\sbgs$ sample yields tighter constraints for $\s8$ and $\omm$. Specifically, it improves $\s8$ constraints by 40\% and $\omm$ constraints by 60\%.} 
    \label{fig:corner_bgs}
\end{figure}

Figure \ref{fig:corner_nbar} shows that including $\nbar$ as a third observable largely improves cosmological constraints when using high-mass galaxy samples with number densities similar to those from cluster studies. However, Figure \ref{fig:corner_nbar} was limited to high-mass galaxies with low number densities. In contrast, lensing and clustering studies using BOSS have often used the full BOSS samples \citep{Leauthaud2017_MNRAS_467_3024, singh2020, sunayama_2020S, Lange2022_MNRAS_509_1779, troster_22} which are an order of magnitude larger in count than complete samples. Lensing and clustering studies with photometric red galaxies typically have higher number densities \citep{desi22}. This raises the question of whether or not it is more advantageous to use low-number density samples that are complete or higher number density samples which will afford better signal to noise on $\ds$ and $\wpp$ but which sacrifice $\nbar$. We now study this question using as our baseline $\sbgs$, a single-bin DESI-like BGS sample consisting of many more galaxies than our $\sfull$ and $\sbin$ samples. Concretely, $\sbgs$ has a number density of $\nbar = 1.70\times 10^{-3}h^3/\mathrm{Mpc}^3$, while $\sfull$ has a number density of $\nbar = 2.36\cdot10^{-5}h^3/\mathrm{Mpc}^3$. Figure \ref{fig:corner_bgs} compares constraints from $\sbgs$ using $\ds$ and $\wpp$ to constraints from $\sbin$ using 
$\ds$, $\wpp$, and $\nbar$. We find that adding $\nbar$ does not compensate for the high signal-to-noise of $\sbgs$. Specifically, the $\sbgs$ sample yields $\omm$ constraints that are 60\% tighter than constraints from $\sbin$, and $\s8$ constraints that are 40\% tighter than constraints from $\sbin$. 

While incomplete samples at face value offer stronger statistical constraining power, it is essential to remember that Figure \ref{fig:corner_bgs} only compares the statistical constraining power of the two approaches. Systematic errors will play an important role in determining the relative merits of both approaches. For DESI, we do not expect to measure  $\m50100$ for the full  BGS sample. Indeed, we would have to adopt a different method of measuring outskirt mass for low-mass galaxies, and it is not clear if outskirt mass would still be a good proxy for low-mass galaxies. The model we have used here is also not optimized for the full BGS sample. For this, we would have to model the galaxy halo connection with a break in the SHMR as we extend the sample to lower masses. In the context of this work, $\sbgs$ is a comparison sample that aims to replicate samples that apply the traditional lensing+clustering cosmology \citep[e.g.][]{Uitert_18, Joudaki_2018, DESY1, desi_19, singh2020, heymans_21, Miyatake2021_arXiv_2111_2419, desi22, amon22, lange_2023}, which would give us a chance to understand the {\it statistical} importance of completeness on cosmological constraints. In this regard, smaller but mass-complete samples may offer significant advantages, particularly concerning modeling systematics (galaxy halo connection, baryonic effects, assembly bias, satellites, etc.). This will be discussed further in Section \ref{dis:lrg}.

\subsection{Impact of satellite galaxies on parameter constraints} \label{sec:satellites}

A key difference between traditional cluster cosmology and super-massive galaxy cosmology is that samples binned by galaxy mass may be contaminated with satellites at the $\sim$10\% level \citep[][etc.]{reid_16, saito_16}. There are multiple ways in which this can be avoided. For instance, one could improve the selection technique of massive galaxy samples to minimize satellite contamination. Alternatively, one could include satellites in the modeling and marginalize over the satellite contribution. However, to begin with, and as a worst-case scenario, here we consider the bias introduced if satellites are ignored altogether. 

We use a mock galaxy catalog created from a snapshot of MDPL2. This mock catalog was created using a subhalo abundance matching \citep[SHAM,][]{vale_04} type approach that matches the HSC stellar mass function derived in \citealt[][]{huang2021outer}. DeMartino et al., {\it in prep} will describe this mock catalog in more detail. We use this mock galaxy catalog simply to extract semi-realistic $\Delta\Sigma$ and $w_{\rm p}$ profiles for satellite galaxies. We then vary the satellite fraction and evaluate the impact on our cosmological fitting.

Previous work has shown that the satellite fraction among galaxies in this mass range ranges between 8-20\%. For example, \citet{reid_16} find a satellite fraction of ~10\%, while \citet{saito_16} find a satellite fraction of ~11\%. We roughly expect galaxies in the S$_{\rm M*, bin}$ samples to have a ~10\% satellite fraction (DeMartino et al., {\it in prep}). Here, we choose to study the effect of satellites in our analysis with an upper limit of ~20\% on the satellite fraction. 

\begin{figure*}
    \centering
    \includegraphics[width=0.9\textwidth]{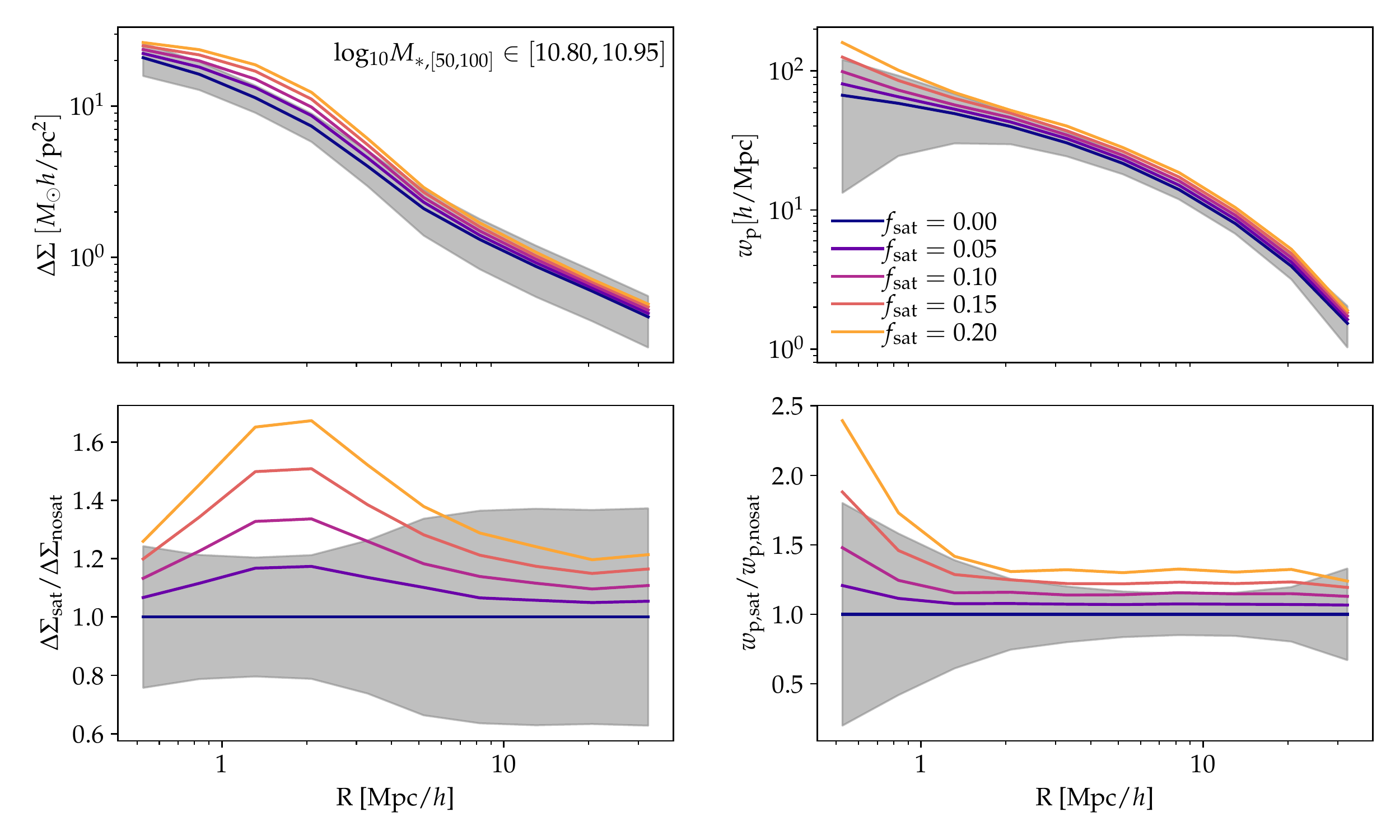}
    \caption{Changes in the shape of  $\ds$ (left) and $\wpp$ (right) due to satellite contamination. Higher satellite fraction impact $\ds$ in the transition between the one and two halo terms, while $\wpp$ in the one halo term. This impact becomes more dominant as $f_\mathrm{sat}$ increases both for $\ds$ and $\wpp$. Predicted HSC error bars are overlaid in grey.}
    \label{fig:satellites}
\end{figure*}

Figure \ref{fig:satellites} displays the effect of satellites on the shape of our key data vectors. We increase the satellite fraction in our selection from 0 to 20\%. As the satellite fraction increases, $\ds$ changes mainly in the transition of the one to two halo term. On the other hand, $\wpp$ is primarily affected in the one halo term. 

\begin{figure}
    \centering
    \includegraphics[width=\columnwidth]{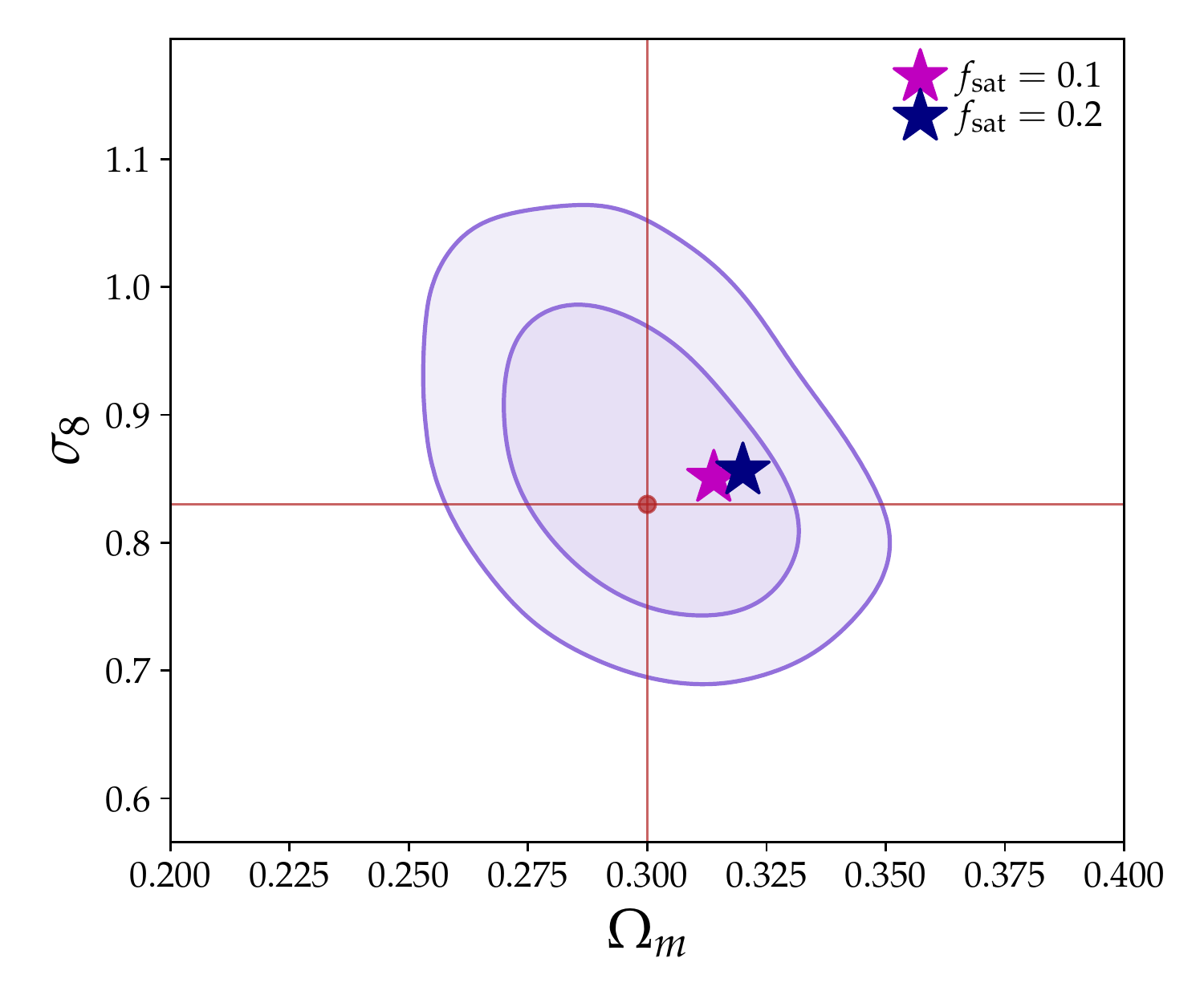}
    \caption{bias in cosmological constraints due to satellite contamination. The posterior contour when $f_\mathrm{sat}=0$ is shown in purple. The maroon dot and lines display the fiducial parameters of the data vector. Two different colored stars represent the mean value of the parameters for $f_{\rm sat}=0.1$ and $f_{\rm sat}=0.2$. Even when satellites are completely ignored, they impart less than a $1\sigma$ shift in cosmological parameters. Therefore, we expect satellites to represent only a minor systematic shift in super-massive galaxy cosmology when properly modeled.} 
    \label{fig:satellites_blobs}
\end{figure}

We create mock data vectors with varying levels of satellite contamination (Figure \ref{fig:satellites}) and analyze these mock data vectors with our fiducial {\cosmolike} model, which does not include a model for satellites. We constrain the model using the same covariance as in the massive galaxies analysis (Figure \ref{fig:corrcof}). The results are shown in Figure \ref{fig:satellites_blobs}. Interestingly, even at the 20\% level, the bias due to satellites is almost negligible compared to our $1\sigma$ cosmological constraints. This may seem surprising given Figure \ref{fig:satellites}. We, therefore, analyze why the cosmology parameters are unaffected.

\begin{figure}
    \centering
    \includegraphics[width=\columnwidth]{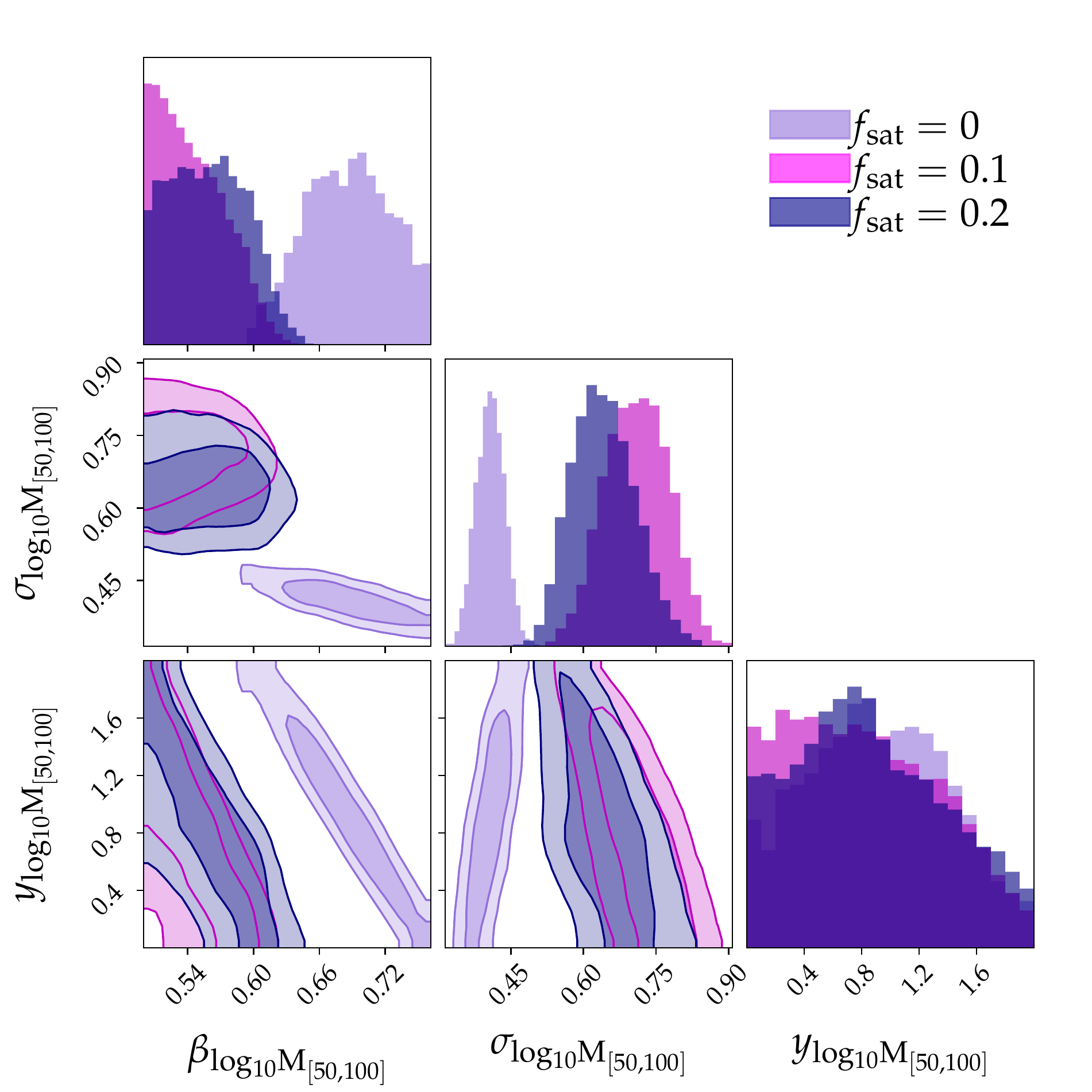}
    \caption{impact of satellites
    on the mass to observable relation parameters. Higher satellite fraction translates into higher $\beta$ and lower $y-$intercept. This strong correlation demonstrates that proper satellite modeling is needed to correctly calibrate the mass to observable relation for massive galaxies.}
    \label{fig:satellites_nuiscance}
\end{figure}

We study how the galaxy-halo parameters are impacted when ignoring satellites. Figure \ref{fig:satellites_nuiscance} displays the constraints on the galaxy-halo parameters when $f_\mathrm{sat} = 0,~0.1~\mathrm{and}~0.2$. Clearly, satellite contamination directly impacts the slope $\beta$ and the $y-$intercept of the mass to observable relation. We conclude that ignoring satellites mainly leads to biases in the galaxy-halo parameters but leaves the cosmology parameters relatively unchanged. In conclusion, while proper satellite modeling is needed to calibrate the mass to observable relation correctly, it is promising to find that satellites do not seem to be a major systematic for super-massive galaxy cosmology.

\section{Discussion}\label{sec:discussion}

\subsection{Massive Galaxies versus Clusters}\label{dis:clusters}

\citetalias{huang2021outer} showed that the outskirt mass of massive galaxies provides a mass proxy that traces halos with comparable scatter to red sequence richness. Given this information, we have demonstrated that the lensing, clustering, and number densities of massive galaxies constrain the growth of structure with similar statistical power to traditional cluster-finding techniques. This is of interest for several reasons. First, massive galaxies could enable us to push to a lower halo mass range where richness becomes uncertain. Second, by pushing down to lower mass, we find that clustering can be used to tighten constraints. Third, we can avoid cluster-finding systematics that are hard to model. Given the statistical precision of current data sets, systematic issues with traditional cluster-finding techniques have become an increasing concern. For example, \citet{desy1miscentering} show that cluster richness estimates tend to be biased low due to miscentering. Furthermore, they indicate that this richness bias may affect cosmology and that future surveys should explicitly take this into account in their cosmological analyses \citep[also see][]{costanzi_19_systematics, sunayama_2020S}. 

Tackling systematics due to selection and projection effects is of primary importance for optical cluster-finding techniques. Massive galaxies present an alternative approach that could potentially bypass some of the more difficult systematics. For example, by selecting halos by their outskirt mass, we avoid the issue of projection effects which have proven difficult to model. This is not to say that this alternative approach will be free of systematics. Additional work will be needed to characterize and understand issues associated with this technique. Here, we discuss three effects that will need to be considered.

The first is satellite contamination and its impact on cosmology. Indeed, we expect that super-massive galaxy samples will have a  $\sim$10\% satellite fraction \citep[e.g.,][]{li_2014, reid_16, saito_16, Leauthaud2017_MNRAS_467_3024, kumar_2022}. However, Figure \ref{fig:satellites_blobs} is promising and suggests that even a 20\% satellite contamination has a less than 1$\sigma$ effect, even with a model that completely ignores satellites. This is because the bias induced by ignoring satellites gets absorbed into the galaxy halo parameters leaving cosmological parameters intact. In addition, in practice, one would attempt to model satellites and not ignore them as we have here. It is plausible that modeling the effects of a 10\% satellite fraction could be more straightforward than modeling red galaxies. This is because the emergence of the red sequence depends on several complex galaxy formation processes that are yet to be well characterized. The second potential systematic is the impact of assembly bias. Indeed, it is possible that galaxies selected by outskirt mass might preferentially select older halos and thus be subject to assembly bias \citep{bradshaw_2020}. This effect is being studied using the TNG simulations and will be presented in upcoming work (Xu et al. {\it in prep}). Finally, we expect baryonic effects \citep[e.g.,][]{Schneider_19, Schneider_20, schneider_20_II, huang_21, shao_22, chen_23} to impact cosmology, especially since we are using both small and large scales in this forecast. However, this systematic is common to both cluster cosmology and super-massive galaxy cosmology, so it needs to be tackled regardless. 

\subsection{Massive Galaxies versus Lensing+Clustering}\label{dis:lrg}

BAO surveys such as BOSS have collected spectra for millions of massive galaxies \citep{dawson_13}. The samples of interest are generally incomplete, except at the highest galaxy masses \citep{leauthaud_16}. This is the main reason why number densities cannot be used as part of the data vector but are instead left to float as a free parameter. The overall number density of a sample is often introduced to scale the normalization of the overall central halo occupation distribution \citep[e.g.][]{Lange2022_MNRAS_509_1779, Yuan2022_MNRAS_tmp_1758}. However,  DESI will be complete in certain mass and redshift range \citep{desi_16}. This allows for restricting analyses to regions of mass and redshift where DESI will be complete. This work analyzes some of the trade-offs in these choices. In Section \ref{sec:lrgs}, we showed the trade-off between using the full DESI BGS sample compared to smaller $M_*$ limited samples. We found that a larger galaxy sample has better constraining power despite not including number density (Figure \ref{fig:corner_bgs}).
 
 However, while at face value, the cosmological constraints from the overall sample outperform the complete samples, our comparisons only consider the statistical constraining power of these different methodologies. When considering systematics, massive $M_*$ limited samples may prove advantageous for several reasons:

 \begin{enumerate}
     \item The galaxy halo modeling of $M_*$ complete super massive galaxy samples is more straightforward than incomplete samples where the effects of color cuts are poorly understood \citep[e.g.,][]{saito_16}. Therefore, $M_*$ complete super massive galaxy samples offer reduced systematic errors associated with the galaxy halo modeling. This may allow smaller radial scales to be used for these samples.
     \item Along similar lines, the impact of assembly bias should be reduced for $M_*$ complete super massive galaxy samples compared to color-selected incomplete samples.
     \item Baryons are expected to modify the galaxy-dark matter connection, the dark matter distribution, and the gas distribution, on scales below a few Mpc. Stellar complete super massive galaxy samples offer a more straightforward way of modeling baryonic effects and their dependence on halo mass \citep[e.g.,][]{Schneider_19, Schneider_20, schneider_20_II} than color-selected incomplete samples. Furthermore, the fact that the proposed galaxy samples have a simple selection function means that comparison with hydrodynamic simulations will be more straightforward.
 \end{enumerate}

 While previous work either uses large scales only or ignores baryonic effects and/or satellite contamination, we expect that mass-complete samples from DESI will be better at simultaneously constraining cosmology, the impact on baryons on the galaxy-dark matter cross-correlation, as well as the halo mass dependence of these effects. Therefore, it could be that the decrease of 30\% in constraining power on $\omm$ (Figure \ref{fig:corner_bgs}) is a worthwhile trade-off for gaining a more simple and accurate galaxy halo modeling as well as constraints on baryonic effects.
 
 Recently, \citet{dvornik_2022} presented constraints on $\s8$ and $\omm$ using the same observational probes as considered here, namely: $\ds$, $\wpp$, and galaxy number densities using the KiDS survey and covering an area of 1006 deg$^2$. \citet{dvornik_2022} use a sample spanning  $0.1<z<1.3$. Their analysis leads to $\s8 = 0.781 ^{+ 0.033} _{-0.029}$ and $\omm = 0.290 ^{+0.021}_{-0.017}$. Here we discuss some differences between this work and our proposed approach. First, \citet{dvornik_2022} adopt photometric samples, whereas we are assuming spectroscopic samples such as those that will be delivered by DESI, which offer high signal-to-noise measurements of $\wpp. $ Furthermore, DESI will avoid all systematics associated with photo-$z$'s (for the lens samples). Second, they use total stellar mass as a proxy which has a larger scatter compared to $\m50100$ \citep{huang2021outer, huang_21}. Third, they push to lower mass than advocated for in this paper. In Figure \ref{fig:satellites_blobs}, we showed that the effect of satellites in the higher mass range of interest is negligible, but this may no longer be the case for low-mass samples with higher satellite fractions. In summary, while  \citet{dvornik_2022} and this paper are similar in spirit, the proposed implementation details are different.

Finally, there is a possibility of using traditional lensing and clustering cosmology jointly with super-massive cosmology. This would utilize the sensitivity that super-massive galaxies have on $\s8$ and $\omm$ in the higher mass range (Figure \ref{fig:change_omm_s8}) and allow the use of $\nbar$ (Figure \ref{fig:corner_nbar}), while also exploiting the high statistical power of the DESI BGS sample (Figure \ref{fig:corner_bgs}). Of course, an accurate understanding of the systematics associated with both samples is needed to reach this point. We will discuss this in more detail in future work.


\section{Summary and Conclusions}\label{conclusions}

In this work, we introduced the idea that super-massive galaxies can be used to trace the growth of structure. We studied how cosmological constraints using massive galaxies as halo tracers compare with those from cluster cosmology and traditional lensing plus projected clustering analyses. We used {\cosmolike} to model our data vector by applying the halo model described in Section \ref{theory}. We build a covariance matrix for the observables $\ds, \wpp, \mathrm{and}~\bar{n}$ assuming 1000 deg$^2$ of HSC lensing data with spectroscopic redshifts from DESI. Our key findings include the following:

\begin{itemize}
    \item We studied the impact of binning and compared how cosmological constraints from three narrow bins in outskirt mass compare to constraints from one cumulative $\m50100$ bin. We found that binning the data in narrow outskirt mass sub-samples improves constraints on $\s8$ by 34\% (Figure \ref{fig:corner_1_v_3}). This is because $\s8$ mainly impacts the high mass slope of the HMF. Because binning helps to constrain this slope, this translates into tighter constraints on $\s8$.
    \item We studied the impact of number density and compared constraints with and without, including $\nbar$ in the data vector. Our results show that including $\nbar$ as a third observational probe improves constraints on $\s8$ by 33\% and $\omm$ by 23\% (Figure \ref{fig:corner_nbar}). 
    \item We compared constraints from a stellar mass limited and mass complete sample to those from a larger but mass incomplete sample (e.g., the DESI BGS). Constraints for a BGS-like sample are tighter in $\omm$ by 60\% and $\s8$ by 40\% (Figure \ref{fig:corner_bgs}). However, we note that these forecasts only consider the statistical constraining power of these different methodologies. We present arguments to suggest that stellar mass limited and mass complete samples may offer distinct advantages when considering the inclusion of systematic effects. 
    \item We study the impact of including projected clustering in our data vector. We find that $\wpp$ strongly impacts our constraining power on $\omm$, representing an 84\% improvement on $\omm$ (Figure \ref{fig:corner_wp}). An analysis using $\ds$ and $\nbar$ but omitting clustering is similar to the approach used in cluster cosmology. While $\wpp$ is not often included as an observational probe in cluster cosmology due to the low number density of clusters, it will be of great utility for super-massive galaxy cosmology where pushing down to lower halo masses may be more straightforward. 
    \item We compare the cosmological constraining power of clusters and massive galaxies as halo tracers. We calibrate a mass to observable relation for each tracer (Figure \ref{fig:mass_to_obs_relations}). Assuming the stellar mass and richness bins in  \citetalias{huang2021outer}, we obtain similar cosmological constraints from both clusters and massive galaxies (Figure \ref{fig:corner_mg_cl}). This results from similar underlying halo mass distributions of our two samples (Figure \ref{fig:halo_mass_dist}). 
    \item One of the main caveats of working with massive galaxies as halo tracers will be satellite contamination. We study this effect and find that for the survey parameters assumed, this is less than a 1$\sigma$ effect on cosmological parameters (Figure \ref{fig:satellites_blobs}). Instead, we find that satellite contamination is absorbed into the parameters of the mass to observable relation (Figure \ref{fig:satellites_nuiscance}). 
\end{itemize}

In this paper, we have shown that massive galaxies present an excellent avenue for performing precision cosmology. Massive galaxies offer competitive constraints to traditional cluster cosmology and will allow us to bypass systematics associated with cluster-finding systematics, such as miscentering and projection effects which can be hard to model and will bias cosmological constraints. This is especially important in the current era of the $S_8$ tension, which could point to new physics or unaccounted systematics. Super-massive galaxies from DESI will be complete down to lower halo masses than $\lambda=20$ cluster samples. By pushing to lower halo masses, $\wpp$ adds strong constraints. There are some caveats to working with massive galaxies. Satellite contamination is inevitable. However, we have shown that cosmological constraints are robust to the impact of satellites. Assembly bias can introduce another potential systematic. Finally, we must consider baryonic effects, which is a systematic for many low redshift probes of the growth of structure. In conclusion, while there is further work to be carried out to turn super-massive galaxies into a full-fledged cosmological probe, this paper has demonstrated that this approach holds tremendous promise because it can push to lower halo masses while simultaneously avoiding systematics associated with cluster finding.

\section*{Acknowledgements}

EX is grateful to Joe DeRose and Tomomi Sunayama for their expert insight on this work. EX acknowledges the generous support of Mr. and Mrs. Levy via the LEVY fellowship. This material is based on AL's work supported by the UD Department of Energy, Office of Science, Office of High Energy Physics under Award Number DE-SC0019301.


\bibliographystyle{mnras}
\bibliography{bibfile}

\label{lastpage}


\end{document}